\documentclass[pra,amsmath,amssymb,floatfix,superscriptaddress,twocolumn]{revtex4-1}

\usepackage{bbm,dsfont,graphicx,mathbbol,mathrsfs,mathtools}
\usepackage{amsmath,graphicx,float,epsfig,float,tabularx,multirow}
\usepackage{epstopdf,color}

\newcommand{\tr}[1]{\mathrm{Tr}\left\{{#1} \right\}} 
 
\newcommand{\ket}[1]{\left\vert {#1} \right\rangle} 
\newcommand{\bra}[1]{\left\langle {#1} \right\vert} 
\newcommand{\eref}[1]{Eq.~(\ref{#1})}

\begin{document}

\title{Out-of-equilibrium Thermodynamics of Quantum Optomechanical Systems}
\author{M. Brunelli}
\affiliation{Centre for Theoretical Atomic, Molecular and Optical Physics, School of Mathematics and Physics, Queen's University Belfast, Belfast BT7\,1NN, United Kingdom}
\author{A. Xuereb}
\affiliation{Centre for Theoretical Atomic, Molecular and Optical Physics, School of Mathematics and Physics, Queen's University Belfast, Belfast BT7\,1NN, United Kingdom}
\affiliation{Department of Physics, University of Malta, Msida MSD 2080, Malta}
\author{A. Ferraro}
\affiliation{Centre for Theoretical Atomic, Molecular and Optical Physics, School of Mathematics and Physics, Queen's University Belfast, Belfast BT7\,1NN, United Kingdom}
\author{G. De Chiara}
\affiliation{Centre for Theoretical Atomic, Molecular and Optical Physics, School of Mathematics and Physics, Queen's University Belfast, Belfast BT7\,1NN, United Kingdom}
\author{N. Kiesel}
\affiliation{Vienna Center for Quantum Science and Technology (VCQ), Faculty of Physics, University of Vienna, 1090 Vienna, Austria}
\author{M. Paternostro}
\affiliation{Centre for Theoretical Atomic, Molecular and Optical Physics, School of Mathematics and Physics, Queen's University Belfast, Belfast BT7\,1NN, United Kingdom}

\begin{abstract}
We address the out-of-equilibrium thermodynamics of an isolated quantum system consisting of a cavity optomechanical device. 
We explore the dynamical response of the system when driven out of equilibrium by a sudden quench of the coupling parameter 
and compute analytically the full distribution of the work generated by the process. We consider linear and quadratic optomechanical 
coupling, where the cavity field is parametrically coupled to either the position or the square of the position of a mechanical oscillator, 
respectively. In the former case we find that the average work generated by the quench is zero, whilst the latter leads to a non-zero 
average value. Through fluctuations theorems we access the most relevant thermodynamical figures of merit, such as the free energy 
difference and the amount of irreversible work generated. We thus provide a full characterization of the out-of-equilibrium thermodynamics 
in the quantum regime for nonlinearly coupled bosonic modes. Our study is the first due step towards the construction and full quantum 
analysis of an optomechanical machine working fully out of equilibrium.
\end{abstract}

\date{\today}

\maketitle

\section{introduction}
As a result of several decades of efforts stemming from different communities, the classical scientific body of thermodynamics have been experiencing  
a true renaissance. The reasons of this revival can mainly be traced back to the release of two constraints: on the one hand the departure from the 
thermodynamic limit, motivated by investigation of increasingly smaller systems, enabled fluctuations to be incorporated; on the other hand the tight 
requirement of quasistatic processes has been relaxed, in favor of generic finite-time transformations connecting 
non-equilibrium states. The overall picture is an exact, non-perturbative extension of thermodynamics to mesoscopic systems lying arbitrarily far from 
equilibrium; stochastic thermodynamics \cite{stochtherm} is now a mature field which addresses thermodynamical quantities such as work, free energy and 
entropy at the level of single trajectories and fluctuation theorems relate the value that these quantities assume at equilibrium to out-of-equilibrium finite-time 
dynamics~\cite{EqIneq, Liphardt}. 
\par  
Furthermore, given the ever-increasing control achievable over microscopic systems and the technological quest for devices miniaturization, one would 
eventually reach a point where quantum fluctuations, besides thermal ones, start playing a non-negligible role \cite{CampisiRev, EspositoRev}. The former 
scenario must then be amended with a full quantum treatment. Performances of thermal machines working in the quantum regime have recently been 
investigated in a plethora of different physical systems~\cite{Mazza}, and the statistics of relevant 
figures of merit such as work and entropy generated  during time-dependent protocols inquired for different models~\cite{Fusco}.
\par
Another motivation to achieve a better understanding of thermodynamics in the quantum regime, somehow complementary with respect to the perspective of
scaling thermal machines down to the nanoscale, comes from the exploration of macroscopic quantum systems. The extension of quantum-limited control over 
objects in the mesoscopic---and possibly macroscopic---domain, is of primary interest both for fundamental problems, e.g. the comprehension of the mechanism of 
decoherence, and for quantum technology. In particular, optomechanical systems provide an ideal platform where to investigate macroscopic quantum phenomena: 
mechanical oscillators made of $10^{15}$ particles are now approaching the quantum regime, offering unprecedented levels of tunability and control \cite{optomacro}. 
For  that reason they are among the most promising candidates to shed light on the interplay between quantum theory and thermodynamics.

\begin{figure*}[t!] 
\centering
\includegraphics[width=2\columnwidth]{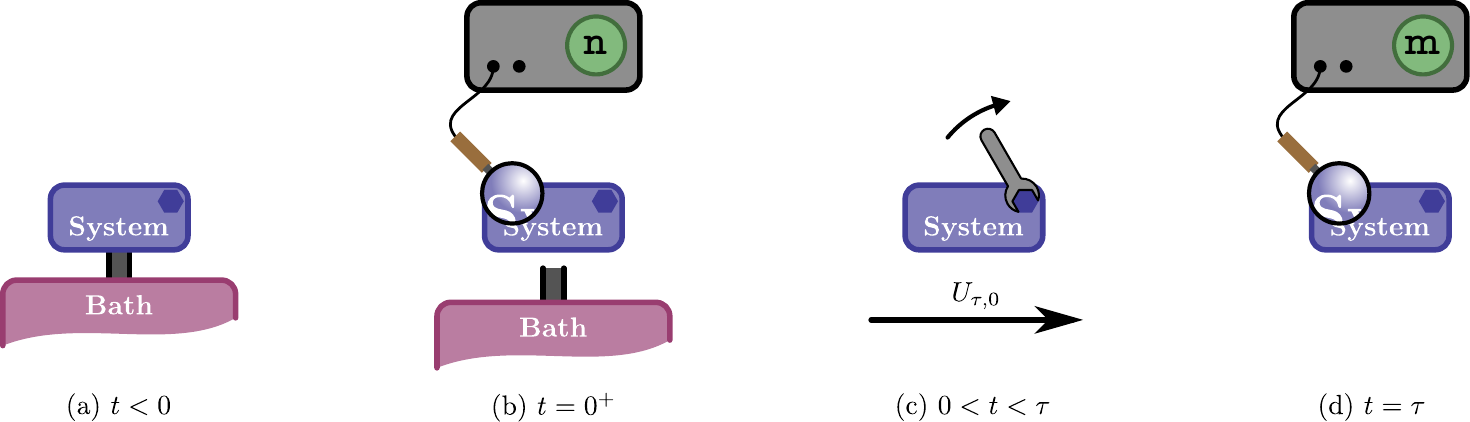}
\caption{Graphical depiction of the two-step protocol for the work distribution. At $t<0$ a system is in contact with a bath until thermal equilibrium is reached [panel (a)]. 
At $t=0^+$, system and bath are detached, while the energy of the system is measured. Let the outcome of such measurement be $E^0_n$, which projects the state of 
the system onto the energy eigenstates $\ket{E^0_n}$ [panel (b)]. The system's Hamiltonian is then changed following to a given protocol and the system evolves according 
to the unitary evolution operator $U(\tau,0)$ for a time $\tau$ [panel (c)], at which time it is measured (over the eigenbasis of the new Hamiltonian). Outcome $E^\tau_m$ is 
achieved, which gives the new state $\ket{E^\tau_m}$ [panel (d)]. By repeating this protocol many times a distribution of values $E^\tau_m-E^0_n$ is achieved, which embodies 
the probability distribution of the work done by/on the system as a result of the protocol that has been implemented.}
\label{scheme}
\end{figure*}

\par
In this work we try to merge these scenarios: We explore and characterize the thermodynamical behavior of an optomechanical system driven out of equilibrium by a 
time-dependent transformation. We address an isolated quantum system, consisting of an optical mode confined in a cavity and parametrically coupled to a mechanical 
oscillator, evolving according to a time-dependent Hamiltonian and undergoing a two-step measurement protocol. Specifically, we will be concerned with a sudden quench 
of the interaction, realized by suddenly switching on the coupling between the two, initially uncoupled, modes. We derive analytic expressions for the characteristic function 
of the work distribution and analyze the full statistics of the work generated. Two different interaction Hamiltonians, both of relevance for present quantum technology, will be 
considered. We shall first discuss the more common case where radiation-pressure interaction couples the cavity field to the position of the oscillator, followed by the case of 
a quadratic optomechanical interaction, where the optical field couples to the square of the position of the mechanical resonator. The starting point for most analyses of 
optomechanical devices is a \emph{linearization} of the interaction, where the Hamiltonian is cast into a quadratic form that is more amenable to analysis. Here, we eschew this 
simplification, which is formally valid when the cavity field is strongly driven~\cite{OptoRev}, and address the full nonlinear optomechanical Hamiltonian. We note at this point that 
the thermodynamical properties of the equivalent linearized model were recently explored by some us in Ref.~\cite{Carlisle}. By retaining the full optomechanical coupling, our 
work therefore aims to address the out-of-equilibrium thermodynamical behavior of nonlinearly coupled bosonic modes in the quantum regime, and thus go beyond the results 
reported in literature so far.
\par
The remainder of this work is organized as follows: {In} Sec.~\ref{thermo} we introduce the {two-measurement} protocol {necessary} to extract the work distribution, and review the quantum 
fluctuation relations. Sec.~\ref{model1} contains a detailed analysis of the dynamical features of an optomechanical system subject to a sudden quench of the coupling parameter and 
assesses its thermodynamical behavior, first in the case of linear {optomechanical} coupling and then in the quadratically-coupled case. Finally, in Sec.~\ref{conc} we summarize our findings 
and discuss new perspectives opened up by this work.

\section{Work distribution and Quantum fluctuation theorems}
\label{thermo}
Let us consider a system described by a time-dependent Hamiltonian $\hat H(G_t)$, whose dependence on time is realized via the externally tunable parameter $G_t$. 
This parameter, which we refer to as the \emph{driving parameter}, determines the configuration of the system at any time. Moreover, let us assume that at $t=0$ the system is 
in thermal equilibrium with a bath at inverse temperature $\beta$, and is hence described by the Gibbs state
\begin{equation}
\hat{\varrho}_{\beta}(G_0)=\frac{e^{-\beta \hat H(G_0)}}{\mathcal{Z}(G_0)} \, ,
\end{equation}  
where $\mathcal{Z}(G_0)=\tr{e^{-\beta \hat H(G_0)}}$ is the canonical partition function of the system. 
This system is taken out of equilibrium by applying a chosen transformation that modifies $G_t$ in time. Here we are concerned with the statistics of the work done on or by the 
system when applying such a protocol. We thus proceed as follows (cf. Fig.~\ref{scheme} for a graphical depiction of the the process): At time $t=0^+$ the system is detached 
from the reservoir and a projective energy measurement is performed on the system in the energy eigenbasis of $\hat H(G_0)$, yielding an eigenstate which we label $\ket{E_n^0}$.
The driving parameter is changed according to the aforementioned transformation until a final time $\tau$. During this period, the state of the system evolves as dictated by the 
action of the unitary evolution operator $\hat U_{\tau,0}$ on the post-measurement state. Finally, a second projective energy measurement is made on the system, this time in the
eigenbasis of $\hat H(G_\tau)$ and yielding eigenstate $\ket{E_m^{\tau}}$. Given the spectral decompositions of the initial and final Hamiltonians, $\hat H(G_0)=\sum_n 
E_n^0\ket{E_n^0}\bra{E_n^0}$ and $\hat H(G_{\tau})=\sum_m E_m^{\tau}\ket{E_m^{\tau}}\bra{E_m^{\tau}}$, respectively, the energy difference between the two outcomes 
$E_m^{\tau}-E_n^0$ may be interpreted as the work performed by the external driving in a single realization of the protocol. This particular value of the work occurs with probability 
$p_n^0 p_{m|n}^{\tau}$, where $p_n^0=e^{-\beta E_n^0}/\mathcal{Z}(G_0)$ keeps track of the initial thermal statistics, while $p_{m|n}^{\tau}=|\bra{E_m^{\tau}}\hat U_{\tau,0}
\ket{E_n^0}|^2$ embodies the transition probability arising from the change of basis. The work performed due to the protocol described above can be characterized by a stochastic 
variable $W$ following the probability distribution  
\begin{equation}\label{pdf}
P(W)=\sum_{n}\sum_{m}p_n^0 p_{m|n}^{\tau}\delta[W-(E_m^{\tau}-E_n^0)] \, .
\end{equation}
Instead of dealing directly with \eref{pdf}, it is often useful  to work with its Fourier transform $\chi(u,\tau)=\int dW e^{\frac{i}{\hbar}uW}P(W)$, which is referred to as the characteristic 
function of the work distribution and can be cast in the form 
\begin{equation}
\label{chitheo}
\chi(u,\tau)=\tr{\hat U^{\dagger}_{\tau,0}e^{\tfrac{i}{\hbar}u\hat H(G_{\tau})}\hat U_{\tau,0}e^{-\tfrac{i}{\hbar}u\hat H(G_0)}\hat{\varrho}_{\beta}(G_0)} \, .
\end{equation}
The utility of the characteristic function becomes apparent when calculating the moments of the work probability distribution explicitly. Indeed, the $k^{\rm th}$ moment of $P(W)$ can 
be obtained from the characteristic function as  
\begin{equation}\label{avwork}
\langle W^k \rangle = (-i\,\hbar)^k \left. {\partial^k_u}\chi(u) \right\vert_{u=0} \, .
\end{equation}  
For the special cases of $k=1,2$ it can be shown that this relation acquires the simple form
\begin{equation}\label{avwork12}
\langle W^k \rangle = \tr{\bigl[\hat H(G_\tau)-\hat H(G_0)\bigr]^k\hat{\varrho}_\beta(G_0)} \, .
\end{equation}  
In what follows we will be concerned with a specific driving protocol, known as sudden quench, where $G_t$ is abruptly changed from its initial value to the final one. 
In this case, $\hat U_{\tau,0}=\mathbb{1}$ and any dependence on $\tau$ disappears. We will thus refer to the characteristic function simply as $\chi(u)$.
\par
Work fluctuation theorems relate the probability distribution of a given process [cf. \eref{pdf}] with its time-reversed counterpart, and account for the emergence of 
irreversibility in isolated systems. In the time-reversed (or backward) process the system is initially in a Gibbs state of the final Hamiltonian $\hat H(G_{\tau})$, and 
the transformation acting on the driving parameter is reversed in time as $G_t\to G_{\tau-t}$. Expressed in terms of the characteristic functions for the forward 
[$\chi(u)$] and backward [$\tilde\chi(u)$] processes, the Tasaki--Crooks fluctuation relation~\cite{crooks} reads 
\begin{equation}\label{tasaki}
\Delta F=\frac{1}{\beta}\ln\left[\frac{\chi(u)}{\tilde{\chi}(i\beta\hbar-u)} \right] \, ,
\end{equation}
where $\Delta F=-\beta^{-1} \log [\mathcal{Z}(G_{\tau})/\mathcal{Z}(G_0)]$ is the free energy difference between the initial states for the forward and backward processes. 
The main implication of this relation is that the probability to extract an amount of work $W$ from the system during the backward process is exponentially suppressed 
with respect to the probability that the same amount of work is done on the system during the forward process. 
\par
Linked to such relation is the celebrated Jarzynski equality~\cite{Jarzynski}
\begin{equation}
\label{jarzynski}
\chi(i \beta\hbar)=\left\langle e^{-\beta W} \right\rangle = e^{-\beta \Delta F} \, ,
\end{equation}   
which links the average of a quantity arbitrarily far from equilibrium with the state function $\Delta F$. From Eq.~\eqref{jarzynski} 
$\Delta F \le \langle W \rangle$ follows immediately, which embodies a statement of the second principle of thermodynamics. The difference between the two quantities, which 
we denote by $W _{\mathrm{irr}}\equiv\langle W \rangle-\Delta F$, is referred to as the irreversible work generated during the transformation.

\section{Work distribution of quenched optomechanical systems}  
\label{model1}
Let us consider the optomechanical interaction between a field mode within a single-mode electromagnetic cavity of resonance frequency $\omega_\mathrm{c}$ and a 
mechanical resonator characterized by its mass $M$ and oscillation frequency $\omega_\mathrm{m}$. These two subsystems will be associated to bosonic annihilation 
operators, denoted by $\hat{a}$ $([\hat{a},\hat{a}^{\dagger}]=\mathbb{1})$ and $\hat{b}$ $([\hat{b},\hat{b}^{\dagger}]=\mathbb{1})$, respectively. The cavity frequency is 
modulated by, and couples parametrically to, the mechanical displacement $x$, so that it can be expanded as  
\begin{equation}
\omega_\mathrm{c}(x)=\omega_\mathrm{c}(0) + x\partial_x \omega_\mathrm{c}(x)\rvert_{x=0}+\tfrac{1}{2}x^2\partial_x^2 \omega_\mathrm{c}(x)\rvert_
{x=0} + {\cal O}(x^2).
\end{equation}

If the leading term in the expansion is the linear one, the two oscillators interact via radiation-pressure and the much-studied linear optomechanical regime is recovered. 
On the contrary, if this term vanishes only the position-squared term contributes so that the so-called quadratic optomechanical regime is accessed; examples of physical 
systems where the latter coupling is achievable are ``membrane-in-the-middle'' setup \cite{quadratic1}, levitating nano-beads \cite{kiesel, nanobead1}, trapped ions or atoms 
\cite{quadratic2}. Note that the adjectives `linear' and `quadratic' here refer to 
the power of the mechanical displacement coupled to the field; we stress, however, that the interaction is inherently nonlinear in the field modes, involving three- or four-wave 
mixing processes. In order to proceed, we assume to be able to control the optomechanical coupling strength, and suddenly turn it on at $t=0^+$. As a function of the 
mechanical position and momentum variables $\hat{x}=x_\mathrm{zpf}\bigl(\hat{b}+\hat{b}^{\dagger}\bigr)$ and $\hat{p}=i(\hbar/2x_\mathrm{zpf})\bigl( \hat{b}^{\dagger}-
\hat{b}\bigr)$, with $x_\mathrm{zpf}=\sqrt{\hbar/2 M \omega_\mathrm{m}}$ the extent of oscillator ground state, the time-dependent Hamiltonian $\hat{H}_t=\hat{H}(G_t)$ reads 
$(t>0)$
\begin{equation}\label{Hgen}
\hat{H}_t=\hbar \omega_\mathrm{c} \hat{a}^{\dagger}\hat{a} + \tfrac{\hat{p}^2}{2M} + \tfrac{1}{2} M \omega_\mathrm{m}^2 \hat{x}^2  + \hbar \,G_t^{(k)}\,
\hat{a}^{\dagger}\hat{a}\,\hat{x}^k \, ,
\end{equation}
where $k=1$ leads to the linear regime and $k=2$ to the quadratic one, $G_t^{(k)}=\Theta(t)k^{-1}\partial_x^k\omega_\mathrm{c}(x)\rvert_{x=0}$ is the coupling parameter, 
and $\Theta(t)$ is the Heaviside step function. Since we set $G_0=0$, both systems are initially uncorrelated and prepared in a global thermal state at inverse temperature $\beta$, 
i.e., $\hat{\varrho}_\beta(G_0)=\hat{\varrho}_{\beta}^\mathrm{(c)}\otimes \hat{\varrho}_{\beta}^\mathrm{(m)}$, where $\hat{\varrho}_{\beta}^{(\alpha)}=\sum_n p_n^{(\alpha)}\ket{n}_
{\!\alpha}\prescript{}{\alpha\!}{\bra{n}}$, with $p_n^{(\alpha)}=N_\alpha^n/(1+N_\alpha)^{n+1}$, and $N_\alpha=(e^{\beta \hbar \omega_\alpha }-1)^{-1}$ being the average number 
of thermal excitations in mode $\alpha=\mathrm{c},\mathrm{m}$. Our main goal is to evaluate the characteristic function of the work distribution \eref{chitheo}, which encompasses 
all the thermodynamically relevant information. Using the above notation, we have
\begin{equation}\label{chiu}
\chi(u)=\tr{e^{\tfrac{i}{\hbar}\hat H_{t>0} u}\,e^{-\tfrac{i}{\hbar}\hat H_0 u}\hat{\varrho}_\beta^{(\mathrm{c})} \otimes \hat{\varrho}_\beta^{(\mathrm{m})}} \, .
\end{equation}
\par
Before moving to the calculation of $\chi(u)$, $P(W)$, and $\Delta F$ for both linear and quadratic coupling cases, let us make a remark about the implementation of 
the quench. The somehow contrasting requirements of having an initial equilibrium state of the cavity--mirror system and turning on the optomechanical interaction at a desired time 
can be reconciled in the following way (here illustrated for the linear coupling case). Let us consider a perfectly reflecting mirror coupled on each side to the field mode
$\hat a_j$ of cavity $\mathrm{c}_j$, $j=1,2$, with equal strength, so that $G_{\mathrm{c}_1}=-G_{\mathrm{c}_2}=G$ and the interaction Hamiltonian will be given by 
$\hat H_{\mathrm{int}}=\hbar G\,(\hat{a}^{\dagger}_1\hat{a}_1-\hat{a}^{\dagger}_2\hat{a}_2)\hat{x}$. If we assume the tripartite system to equilibrate and consider the reduced state 
of one cavity mode and the mirror we have $\hat{\varrho}^{(\mathrm{c_1 m})}= \mathcal{Z}_{\mathrm{c}_1}\mathcal{Z}_{\mathrm{c}_2}\mathcal{Z}_{\mathrm{m}}\mathcal{Z}^
{-1}_{\mathrm{c_1c_2m}}\sum_{n,m}p_n^{(\mathrm{c}_1)}p_m^{(\mathrm{c}_2)}e^{\beta \hbar \omega_\mathrm{m}\mu_{n,m}^2 }\times \\ \hat D^{\dagger}(\mu_{n,m})
\hat{\varrho}_\beta^{(\mathrm{m})}\hat D(\mu_{n,m})\otimes \ket{n}{\bra{n}}_\mathrm{c_1}$ where $\mu_{n,m}=G (x_\mathrm{zpf}\,\omega_\mathrm{m})^{-1}(n-m)$. We can
see that, unless the thermal states of the two cavities are perfectly correlated (in a classical way), this state does not reduce to $\hat{\varrho}_\beta^{(\mathrm{c_1})}
\otimes \hat{\varrho}_\beta^{(\mathrm{m})}$, namely the initial state required by the protocol. However, we computed the Kullback--Leibler divergence of the diagonal part 
$\hat{\varrho}^{(\mathrm{c_1 m})}$ (the only entering the protocol) with respect to thermal statistics $p_n^{(\mathrm{c}_1)}p_k^{(\mathrm{m})}$, and we found that in the 
range of parameters explored in this work it never exceeds values of the order of $10^{-4}$. Therefore, this configuration may provide a viable method for approximating the 
initial state of the protocol. The quench would then consist in the sudden shut-off of the auxiliary mode $\hat a_2$. A detailed feasibility analysis of the whole protocol is
however beyond the scope of this work and it is left for future investigations.

\subsection{Quenched linear optomechanical interaction}
For the case of a Fabry-P\'erot cavity of length $L$ and oscillating mirror of mass $M$ the coupling can be shown to be equal to $G^{(1)}_{t>0}=\omega_\mathrm{c}/L\equiv 
g/x_\mathrm{zpf}$, where $g$ is referred as the single-photon coupling strength and quantifies the shift in the equilibrium position of the mechanical resonator induced by 
a single photon. In order to keep the notation as simple as possible, we will explicitly denote by $\hat H_\mathrm{I}$ the (initial) uncoupled Hamiltonian 
\begin{equation}\label{HI}
\hat H_{t=0}=\hbar \omega_\mathrm{c} \hat{a}^{\dagger}\hat{a}+ \hbar \omega_\mathrm{m}(\hat{b}^{\dagger}\hat{b}+\tfrac{1}{2})\equiv \hat H_\mathrm{I} \, ,
\end{equation}
and by $\hat H_\mathrm{F}$ the {(final)} interacting one
\begin{equation}\label{HF}
\hat H_{t>0}=\hat H_\mathrm{I} + \hbar \,g\,\hat{a}^{\dagger}\hat{a}
(\hat{b}+\hat{b}^{\dagger})\equiv \hat H_\mathrm{F} \, .
\end{equation}%
It is straightforward to prove that 
\begin{equation}
\label{bch}
\begin{aligned}
e^{-\tfrac{i}{\hbar}\hat H_\mathrm{F} u}&=e^{-i\omega_\mathrm{c}u\,\hat{a}^{\dagger}\hat{a}+i\tfrac{g^2}{\omega^2_\mathrm{m}}(\omega_\mathrm{m}u -\sin\omega_
\mathrm{m}u)\,(\hat{a}^{\dagger}\hat{a})^2}\\
&\times e^{-\tfrac{g}{\omega_\mathrm{m}}\hat{a}^{\dagger}\hat{a}(\eta \hat{b}^{\dagger}-\eta^*\hat{b})} e^{-i\omega_\mathrm{m}u\,\hat{b}^{\dagger}\hat{b}}\, ,
\end{aligned}
\end{equation}
where $\eta=(1-e^{-i\omega_\mathrm{m}u})$~\cite{Bose}. Expression~(\ref{bch}) provides us with physical insight into the dynamical evolution induced by radiation-pressure 
interaction: Apart from two free-rotating terms (the first and last in the above product), the propagator reduces to a displacement of the mechanical mode conditioned on the 
number of cavity photons, followed by an evolution generated by a Kerr-like term.
\par
The characteristic function in Eq.~(\ref{chiu}) can then be explicitly worked out. The form of the interaction suggests taking the trace over the number states $\{\ket{n}_\mathrm{c}\}$ for mode 
$\hat{a}$ and over the coherent states $\{\ket{\alpha}_\mathrm{m}\}$ for $\hat{b}$ (we reserve Latin letters for Fock-state labels and Greek letters for coherent-state labels throughout), i.e.,
\begin{equation}
\chi(u)=\sum_{n=0}^{\infty}\int_{\mathbb{C}}\mathrm{d}^2\alpha\, p_n^\mathrm{(c)}\, {\cal P}^\mathrm{(m)}(\alpha)\bra{n,\alpha}
e^{\tfrac{i}{\hbar}\hat H_\mathrm{F} u}e^{-\tfrac{i}{\hbar}\hat H_\mathrm{I} u}\ket{n,\alpha}\,,
\end{equation}%
where ${\cal P}^\mathrm{(m)}(\alpha)=\exp{(-|\alpha|^2/N_\mathrm{m})}/(\pi N_\mathrm{m})$ is the Glauber--Sudarshan $P$-representation of an equilibrium thermal state in the coherent-
state basis and the compound kets are defined as $\ket{n,\alpha}\equiv\ket{n}_\mathrm{c}\otimes\ket{\alpha}_\mathrm{m}$. It is possible to gather the following analytical expression for the
characteristic function
\begin{equation}
\label{chi}
\chi(u)=\sum_{n=0}^{\infty}\frac{N_\mathrm{c}^n e^{-\tfrac{g^2n^2}{\omega^2_\mathrm{m}}[i(\omega_\mathrm{m}u-\sin\omega_\mathrm{m}u)+(1+2 N_\mathrm{m})(1-\cos\omega_
\mathrm{m}u)]}}{(1+N_\mathrm{c})^{n+1}}
\end{equation}
which cannot be summed analytically. We can however appreciate a few significant features of such expression: First, we recognize the thermal statistics of the cavity field modulated by an 
exponential whose argument keeps track of the average number of phonons $N_\mathrm{m}$. Second, the characteristic function is  periodic in $u$. 
\begin{figure}[t] 
\centering
\includegraphics[scale=.28]{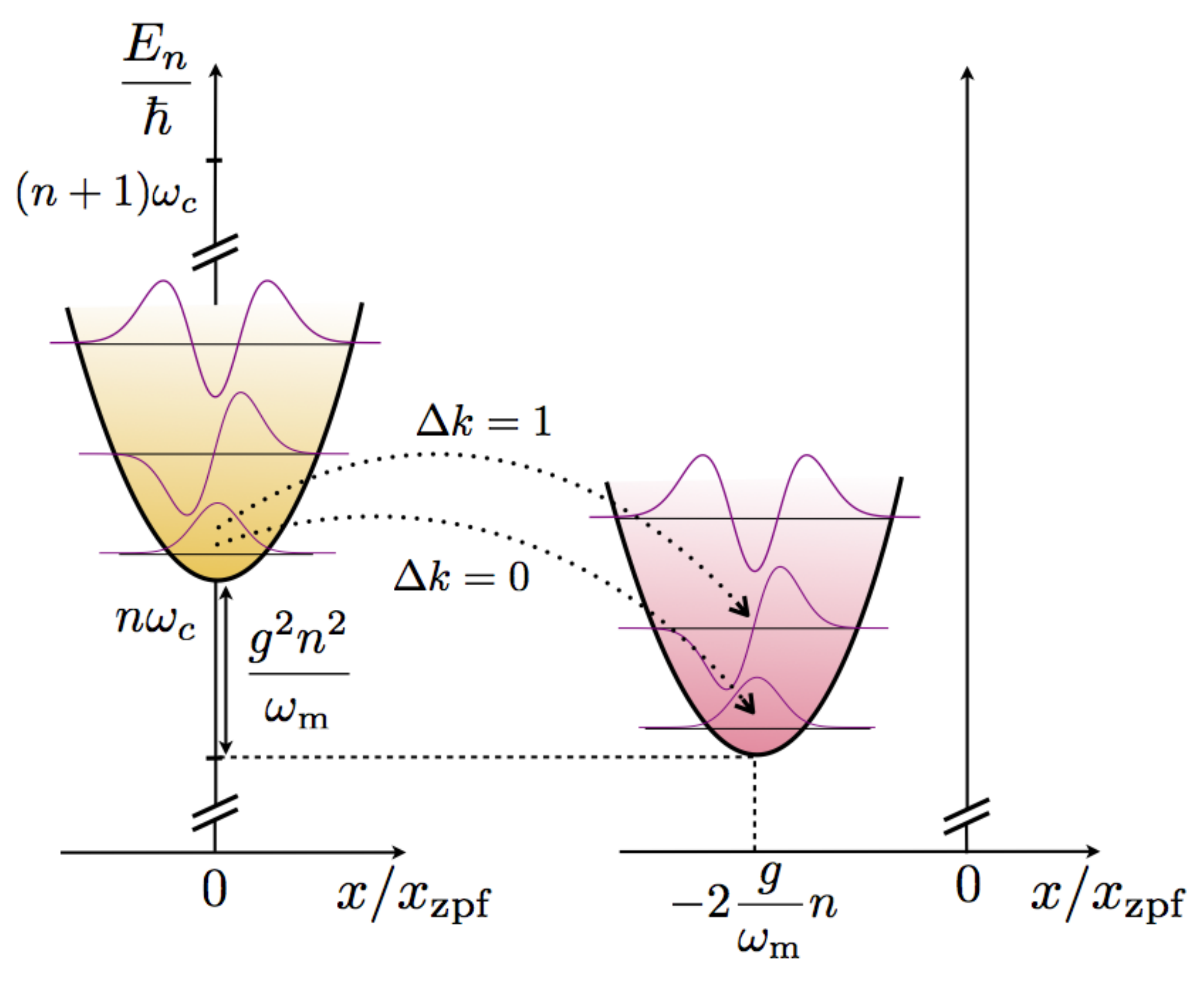}
\caption{Schematic diagram (not to scale) of the energy-level structure of the pre-quench, $\hat H_{\mathrm{I},n}$, and post-quench, $\hat H_{F,n}$, Hamiltonians for the $n$-photon manifold.
Quenching the linear optomechanical interaction results both in an energy shift and a displacement of the machanical oscillator. Two possible transitions induced by the quench---having different 
values of $\Delta k=k'-k$---are shown as an example.}
\label{f:linearquench}
\end{figure}

To proceed further, since the Fourier transform of Eq.~(\ref{chi}) cannot be explicitly worked out, we evaluate the probability distribution of the work by calculating \eref{pdf} directly.
To do this, we require the energy eigenvalues and eigenstates of $\hat H_\mathrm{I}$ and $\hat H_\mathrm{F}$. As $\hat H_\mathrm{I}$ is the free Hamiltonian of the uncoupled system, 
it satisfies the eigenvalue equation $\hat H_\mathrm{I}\ket{n,k}=E_{n,k}\ket{n,k}$, where $\ket{n,k}=\ket{n}_{\rm c}\otimes\ket{k}_{\rm m}$, and $E_{n,k}=\hbar\omega_\mathrm{c}n+
\hbar\omega_\mathrm{m}(k+\tfrac{1}{2})$. Owing to the fact that $[\hat{a}^{\dagger}\hat{a},\hat H_\mathrm{F}]=0$, the post-quench Hamiltonian can be written as $\hat H_\mathrm{F}=
\bigoplus_{n=0}^{\infty}\hat H_{\mathrm{F},n}$, where $\hat H_{\mathrm{F},n}=\ket{n}{\bra{n}}_\mathrm{c}\bigl[\hbar \omega_\mathrm{c} n+ \hbar \omega_\mathrm{m}(\hat{b}^{\dagger}
\hat{b}+\tfrac{1}{2})+ \hbar \,g\,n(\hat{b}+\hat{b}^{\dagger})\bigr]$ refers to the Hamiltonian of the $n$-photon manifold. Each $\hat H_{F,n}$ can then be diagonalized using a displacement 
operator $\hat D(z)=\exp(z \hat{b}^{\dagger}-z^*\hat{b})$ on the mechanical mode, whose amplitude we take conditioned to the photon number $n$~\cite{Girvin}. Denoting the quantities referring to 
$\hat H_{\mathrm{F},n}$ with a prime we find the energy eigenstates, written in the energy eigenbasis of the \emph{initial} Hamiltonian $\hat H_\mathrm{I}$, $\ket{n'}_\mathrm{c}  
\otimes \hat D^{\dagger}(\frac{g \,n'}{\omega_\mathrm{m}}) \ket{k'}_\mathrm{m}$, with eigenvalues $E_{n',k'}=\hbar \omega_\mathrm{c} n' + \hbar \omega_\mathrm{m}(k'+\tfrac{1}{2}) -
\hbar \frac{g^2}{\omega_\mathrm{m}} n'^2$. A pictorial view of pre- and post-quench eigenstates in the subspace at fixed number $n$ of photons is sketched in Fig.~\ref{f:linearquench}. 
As stated by \eref{pdf}, the transitions from a set of eigenstates to another are responsible---at the microscopic level---for the work performed on or by the system. The probability distribution 
of the work is thus given by
\begin{equation}
\begin{aligned}\label{PW}
P(W)&= \sum_{n,n',k,k'}p_n^\mathrm{(c)}p_k^\mathrm{(m)}\lvert\prescript{}{\mathrm{m}}{\bra{k'}}\hat D[(g/\omega_\mathrm{m})n']\ket{k}_\mathrm{m}\rvert^2 \\
&\times\delta[W-(E_{n',k'}-E_{n,k})] \delta_{n,n'}\\
&= \sum_{n,k,k'}p_n^\mathrm{(c)}p_k^\mathrm{(m)} \frac{k!}{k'!}\,e^{-(g/\omega_\mathrm{m})^2n^2}[(g/\omega_\mathrm{m})n]^{2(k'-k)} \\
&\times\Bigl\{\mathscr{L}_{k}^{(k'-k)}[(g/\omega_\mathrm{m})^2n^2]\Bigr\}^{\!2} \\
&\times\delta\{W-\hbar\omega_\mathrm{m}[k'-k-(g/\omega_\mathrm{m})^2n^2]\} \, ,
\end{aligned}%
\end{equation}
where $\mathscr{L}_{a}^{b}(x)$ are the generalized Laguerre polynomials coming from the evaluation of the overlap between pre- and post-quench mechanical oscillator eigenstates
~\cite{dispnumber}. A comparison with Eq.~(\ref{pdf}) enables to unambiguously discriminate the contribution of the first projective measurement (which consist of a sampling from 
the joint thermal distribution of the cavity and the mirror) from the quantum transition probability, and explicitly provides an analytical expression for the latter.  
\begin{figure}[h!] 
\centering
\includegraphics[scale=.46]{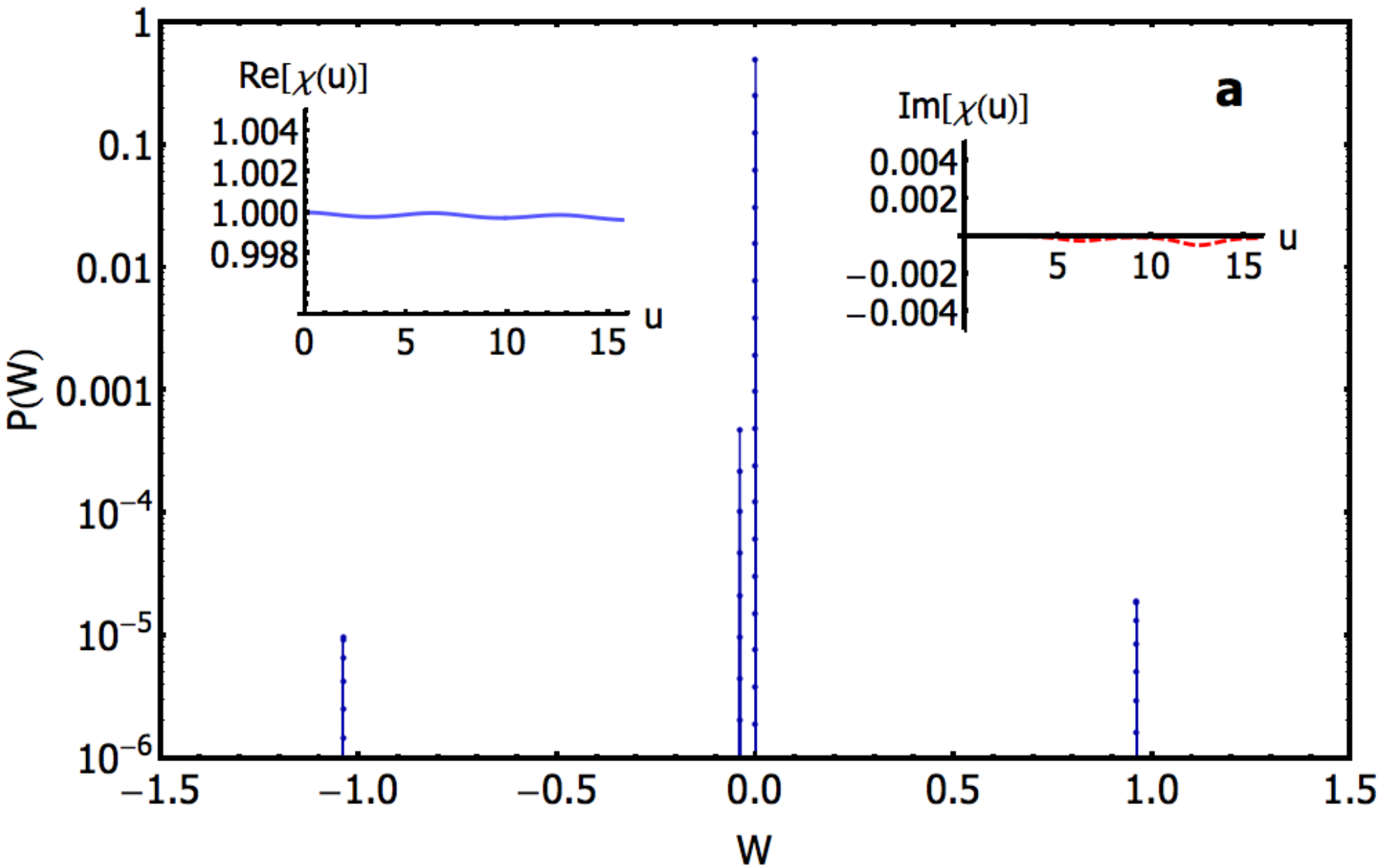}
\\
\includegraphics[scale=.46]{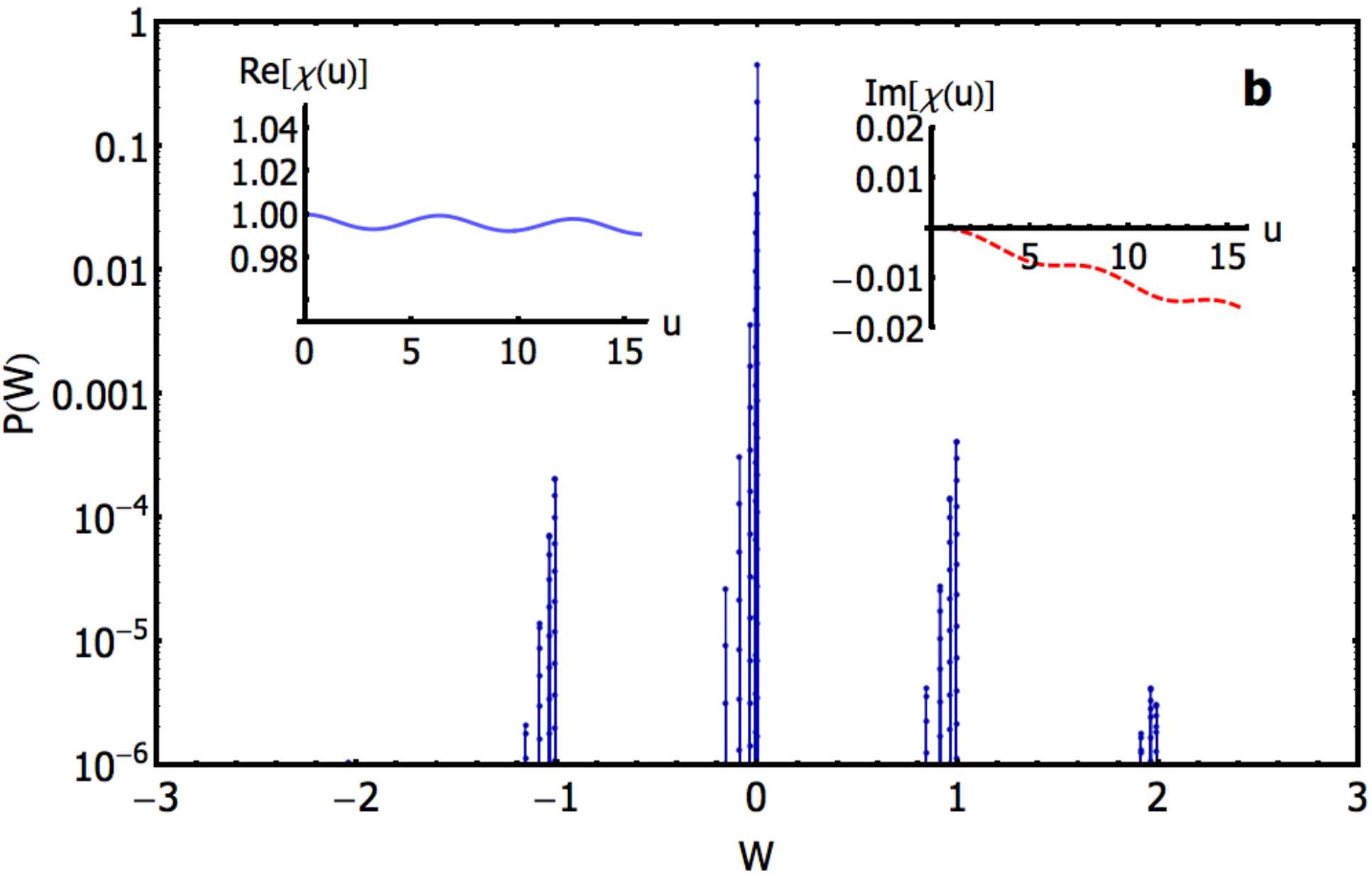}
\\
\includegraphics[scale=.46]{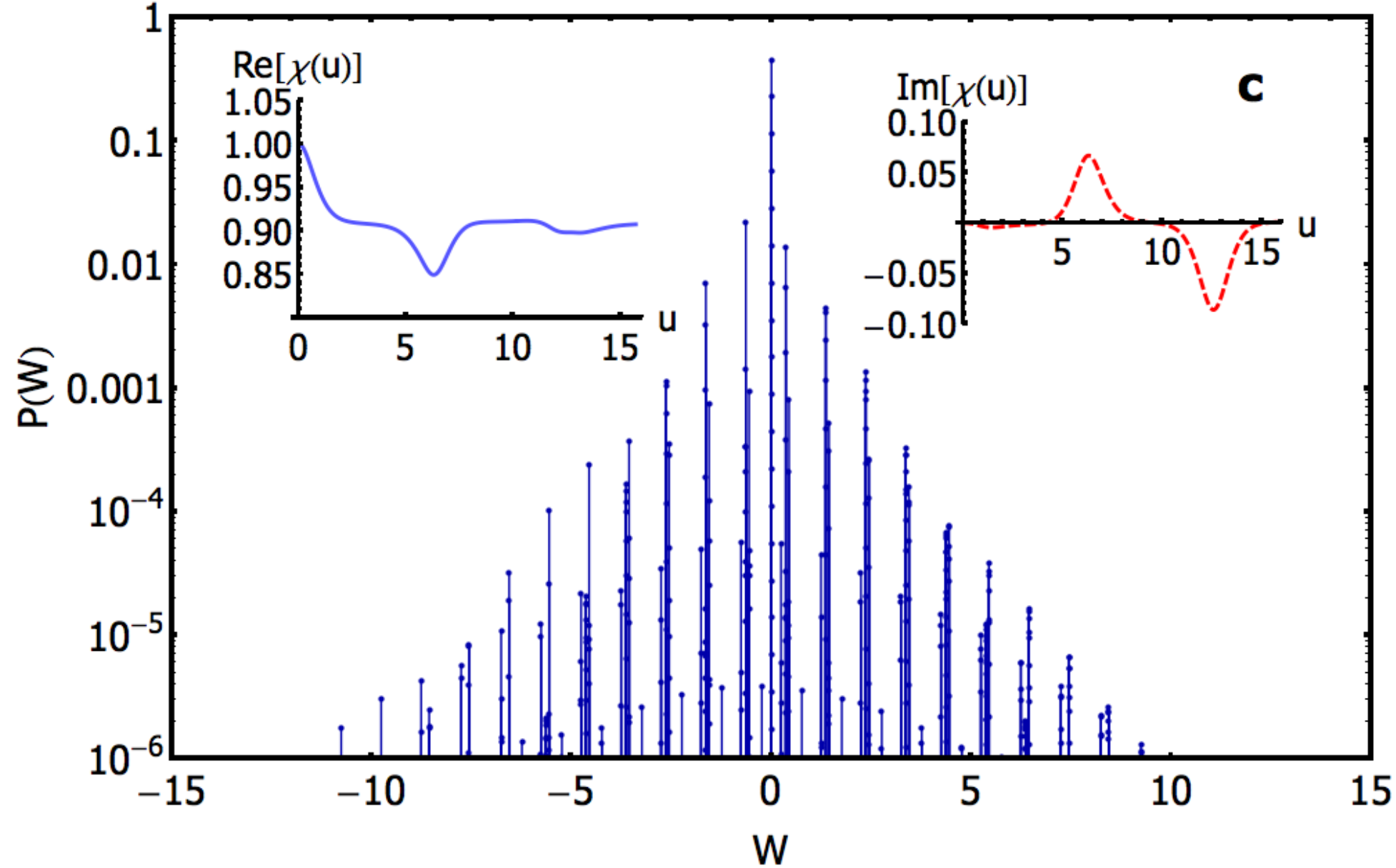} 
\caption{Logarithmic plot of the probability distribution of the stochastic work variable, $W$ (in units of $\hbar \omega_\mathrm{m}$) for different values of the average number of 
cavity photons $N_\mathrm{c}$, average number of mechanical phonons $N_\mathrm{m}$ and coupling $g$. Panel ({\bf a}) is for $(N_{\mathrm{c}}, N_{\rm m}, g)=(0.001,0.1,0.2
\omega_\mathrm{m})$, ({\bf b}) is for $(N_{\mathrm{c}}, N_{\rm m}, g)=(0.1,1,0.1\omega_\mathrm{m})$ while ({\bf c}) for $(N_{\mathrm{c}}, N_{\rm m}, g)=(0.1,1,0.8\omega_\mathrm{m})$.  
In the inset is shown the behavior against the time-like variable $u$ (multiplied by $\omega_\mathrm{m}$) of the real, $\mathrm{Re}(\chi)$ (solid blue, left), and imaginary, 
$\mathrm{Im}(\chi)$ (dashed red, right) parts of the characteristic function.} 
\label{f:Chi}
\end{figure}
The probability distribution of the work, together with real and imaginary parts of the characteristic function, is shown in Fig.~\ref{f:Chi}, for different values of $N_\mathrm{c}$, 
$N_\mathrm{m}$, and coupling strength. By differentiating the expression of characteristic function \eref{chi} and evaluating it in the origin, according to the prescription in 
\eref{avwork}, one can see that each term of the series identically vanishes, so that the average work generated by quenching the optomechanical coupling is in fact zero. 
This is in agreement with the behavior of the imaginary part of $\chi(u)$, shown in the inset of Fig.~\ref{f:Chi}, which approaches $u=0$ with zero derivative; the distribution of 
the work values is therefore centered around $W=0$. Having access to the characteristic function also gives us information about the statistical moments of $P(W)$; e.g., the 
variance of the distribution is given by 
\begin{equation}
\langle W^2\rangle-\langle W\rangle^2=\hbar^2g^2N_\mathrm{c} (1+2N_\mathrm{c})(1+2N_\mathrm{m}) \,.
\end{equation}
As expected, this quantity increases  both with respect to the intensity of the quench, as quantified by $g/\omega_\mathrm{m}$, and the average number of thermal excitations. 
This feature is apparent by comparing the topmost distribution, relative to $N_\mathrm{c}=0.001$, $N_\mathrm{m}=1$ and $g/\omega_\mathrm{m}=0.2$, to the other two, 
both obtained for  $N_\mathrm{c}=0.1$ and $N_\mathrm{m}=1$---thus varying the ratio $\omega_\mathrm{c}/\omega_\mathrm{m}$---but corresponding to $g/\omega_\mathrm{m}=0.1$
and $g/\omega_\mathrm{m}=0.8$ respectively, i.e., increasing both the temperature and the coupling strength.
\par
Let us first analyze $P(W)$ as illustrated for a few representative cases in Fig.~\ref{f:Chi}, where we consider small values of $g/\omega_\mathrm{m}\lesssim1$. In such conditions 
and for relatively small values for $N_\mathrm{c}$, the probability distribution appears to be dominated by peaks occurring close to multiple values of $\hbar\omega_\mathrm{m}$. 
These peaks originate from different initially-populated Fock states of the mechanical subsystem. Indeed, the number of peaks with appreciable amplitude increases strongly with 
$N_\mathrm{m}$. In Fig.~\ref{f:Chi} ${\bf (b)}$ we notice that the sparse peak-distribution associated with very low values of $N_{\rm c}$ changes into a ``clustered" one, where 
groups of peaks develop close to multiples of $\hbar\omega_{\rm m}$ and are biased towards less positive values of $W$. This is directly caused by the Kerr-like term in 
$\hat H_\mathrm{F}$, whose contribution to the overall energy is always negative. A natural question to ask at this point is why the average work done is zero when each of these 
fine structures is biased in the same direction. The answer to this lies in the positive skewness of the distribution, which is given by  
\begin{equation}
\gamma=\frac{\langle(W-\langle W\rangle)^3 \rangle}{\langle(W-\langle W\rangle)^2 \rangle^{3/2}}=\frac{\omega_\mathrm{m}/g}{(1+2N_\mathrm{m})\sqrt{N_\mathrm{c} (1+2N_\mathrm{c})}} \,,
\end{equation}
and is more apparent in the low-temperature regime; indeed, by simply looking at the distribution shown in Fig.~\ref{f:Chi} ({\bf b}), it is possible to appreciate the positive skewness of the 
distribution.
\begin{figure}[t] 
\centering 
\includegraphics[scale=.46]{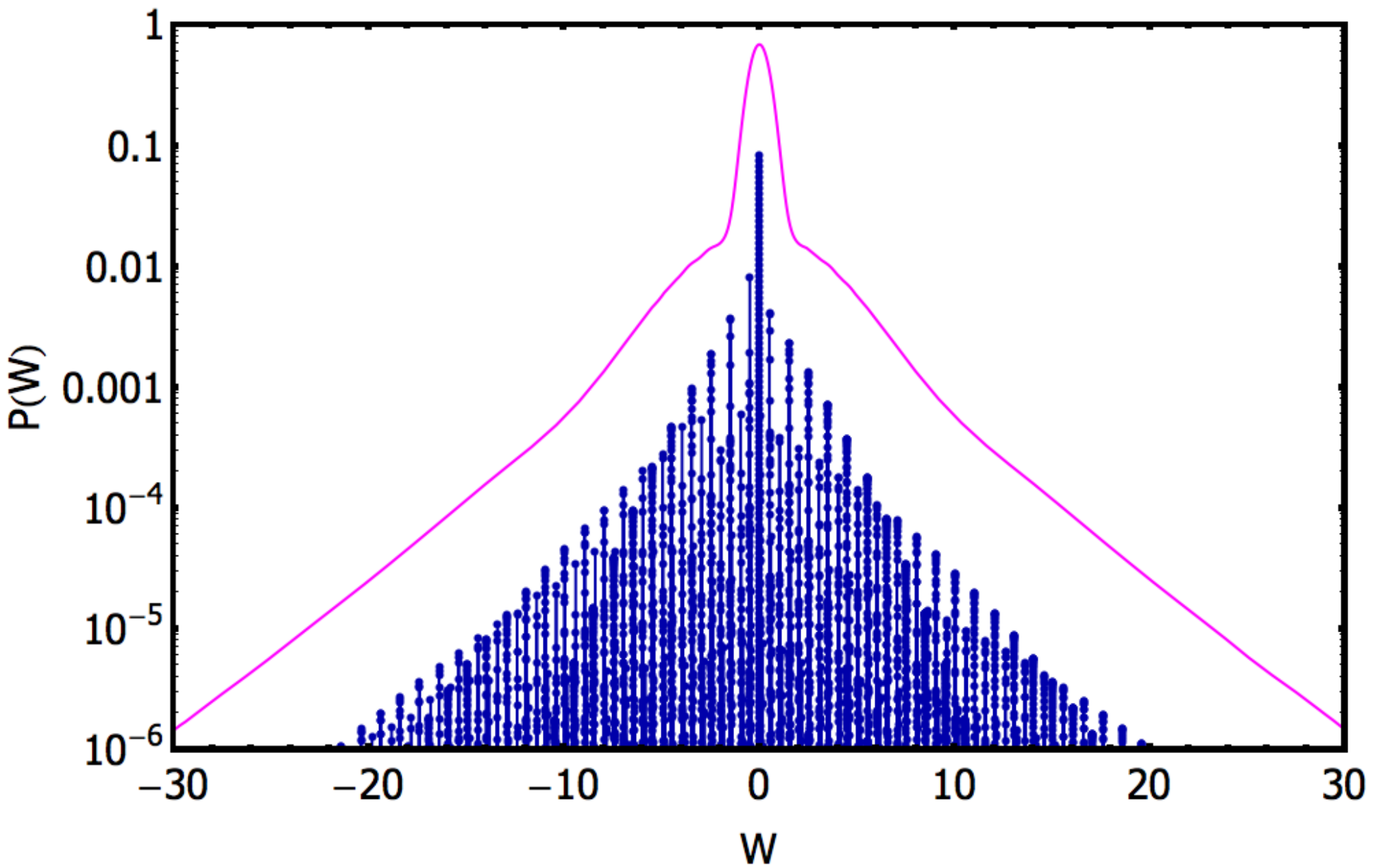}
\\
\includegraphics[scale=.46]{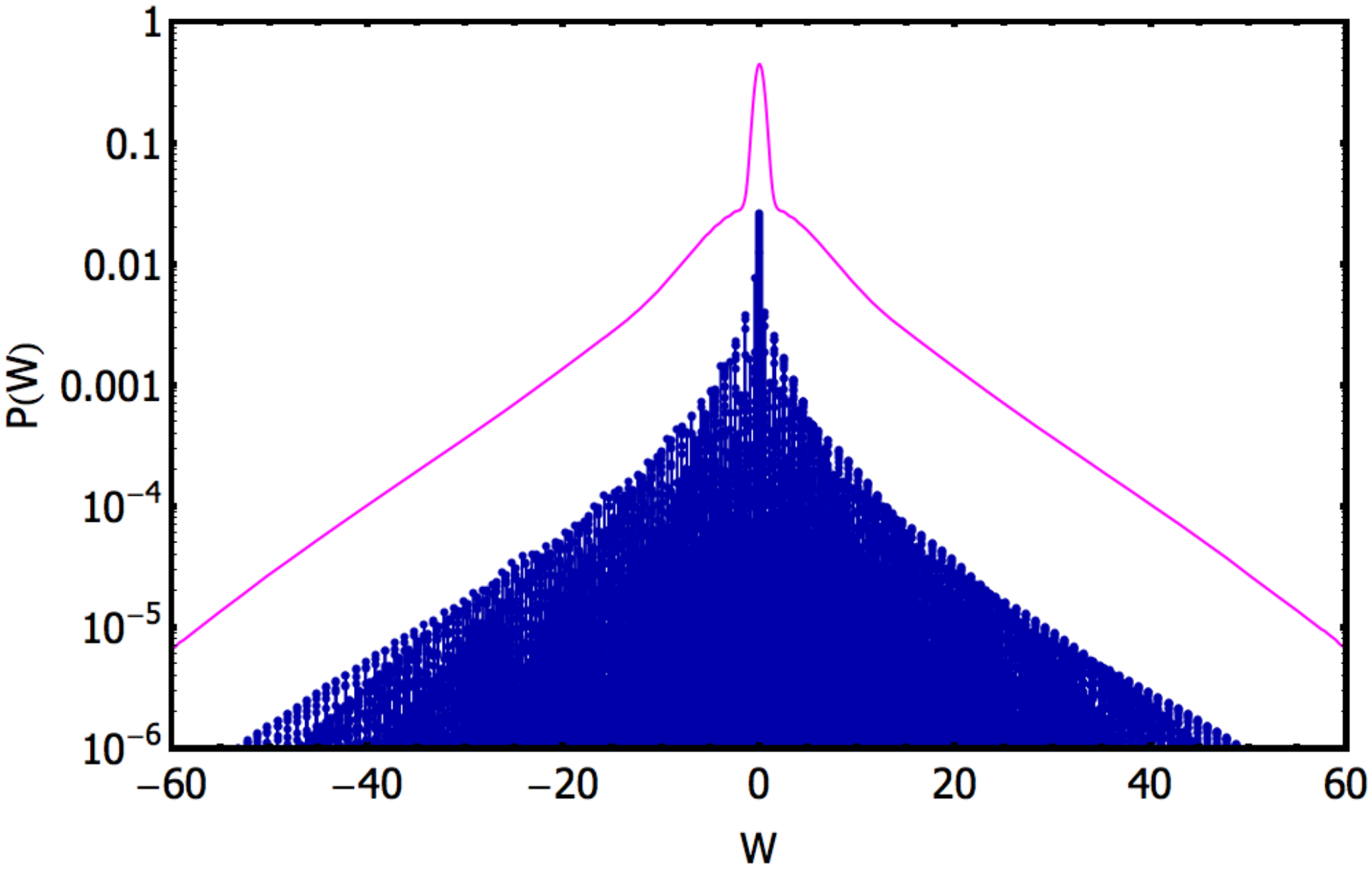}
\caption{Logarithmic plot of the probability distribution of the work (in units of $\hbar \omega_\mathrm{m}$) corresponding to the parameters $(N_c, N_{\rm m}, g)=
(0.19,9, 0.7\omega_{\rm m})$ [$(N_c, N_{\rm m}, g)=(0.9,19, 0.7\omega_{\rm m})$]  for the upper panel [for the lower panel]. The solid magenta line shows the 
coarse-grained version of the distribution. \label{f:CGlinear}}
\end{figure}
\par
Shifting our attention from Fig.~\ref{f:Chi} to Fig.~\ref{f:CGlinear}, we can appreciate the effects of increasing the temperature significantly. The two effects we discussed above, 
namely the increasing number of peaks upon increasing $N_\mathrm{m}$ and the fine structure that appears more and more prominently when increasing $N_\mathrm{c}$, work 
together to turn $P(W)$ from a distribution consisting of well-separated peaks to a dense forest of points. It is readily apparent from the latter figure that the tails of the distribution 
decay exponentially with increasing $\lvert W\rvert$. In order to investigate this effect more thoroughly, we show in Fig.~\ref{f:CGlinear} a coarse-graining of the probability distributions. 
This coarse-graining was performed by convolving $P(W)$ with a Gaussian of appropriate width ($0.5\,\hbar\omega_\mathrm{m}$ in this case). The resulting distributions, drawn as solid 
curves in this figure, display clearly a tripartite structure. First, around $W=0$, a prominent peak is apparent whose width in this figure is entirely due to the convolved Gaussian. Second, 
a quadratic decay is appreciated for slightly larger values of $W$. The probability distribution in this region is thus Gaussian in nature. Third, the tails of the distribution have a manifestly 
exponential character: the coarse-grained curve displays a prominent kink where the exponential tail meets the Gaussian part of the distribution. 
\par
It is worth discussing the validity of our coarse-graining approach. We have verified that the discussion above is not modified significantly when the function used to coarse-grain is changed 
from a Gaussian or a Lorentzian, or when the width of this function is changed within reason. A final check we performed was to construct the cumulative distribution function  
$\int_{-\infty}^W\mathrm{d}w\,P(w)$. This function was interpolated and smoothed, and then differentiated to give a continuous version of $P(W)$. Once again, the conclusions we drew above 
were left unmodified. It is possible to attach a physical meaning to the coarse-graining of $P(W)$ as follows. Should the probability distribution be measured using any realistic apparatus, the 
measurement results will not be infinitely sharp, and will be distributed according to some distribution, usually assumed to be Gaussian. Such an experiment would directly yield the coarse-grained 
distribution we calculate and display in Fig.~\ref{f:CGlinear}.
\par
We have thus shown, analytically and numerically, that despite turning on a nonlinear interaction between the two modes, on average there is no net production of work. This is perhaps a 
surprising fact, given that it has been established that either by quenching the frequency of the harmonic potential of a single oscillator~\cite{Galve}, or the linear interaction between two 
bosonic modes~\cite{Carlisle}, net work \emph{is} produced on average. We shall return to this point in the next subsection, where we discuss the physical origin of this fact and demonstrate 
a method for producing non-zero average work.
\begin{figure}[t] 
\centering 
\includegraphics[scale=.21]{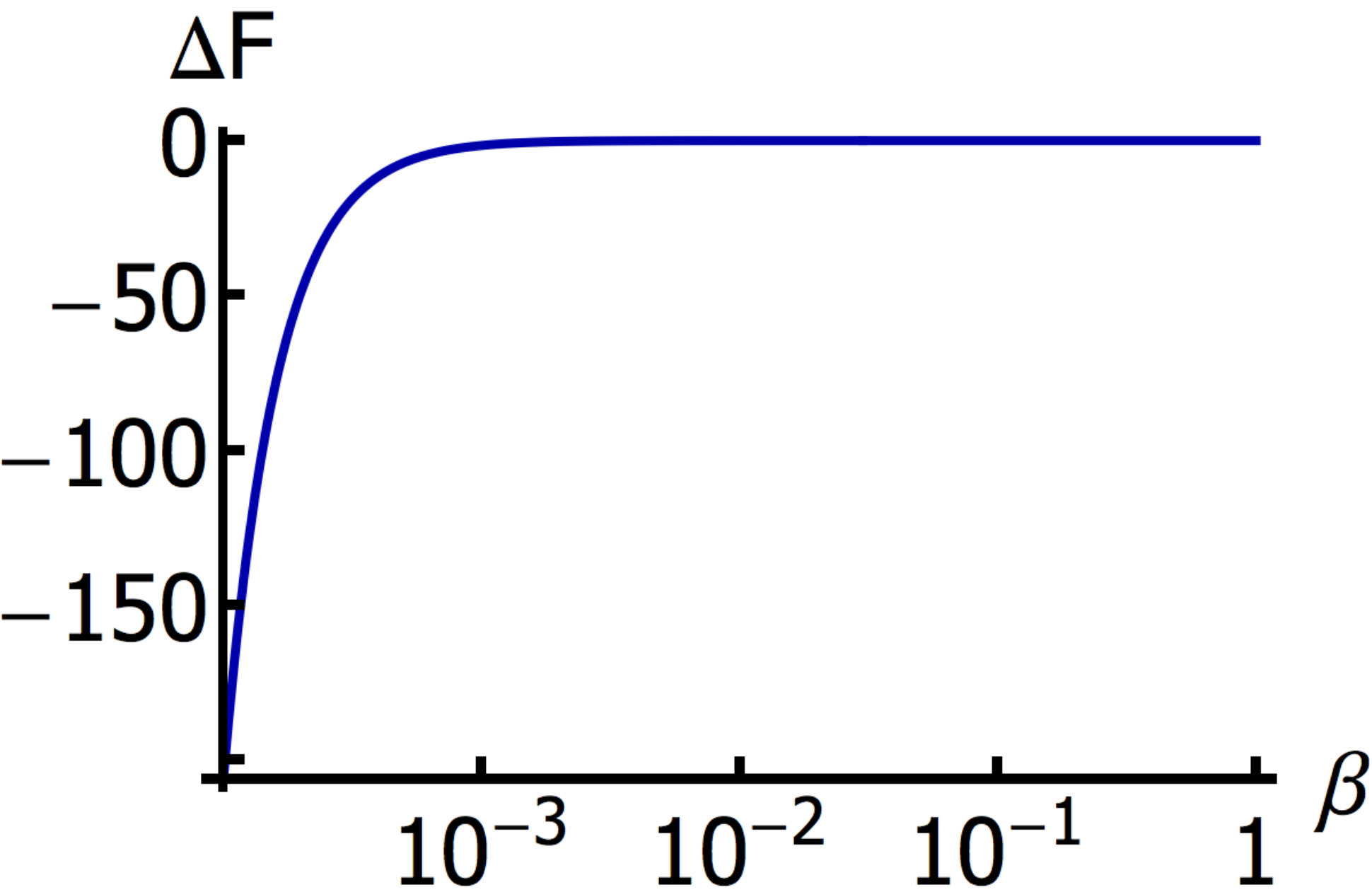}
\includegraphics[scale=.21]{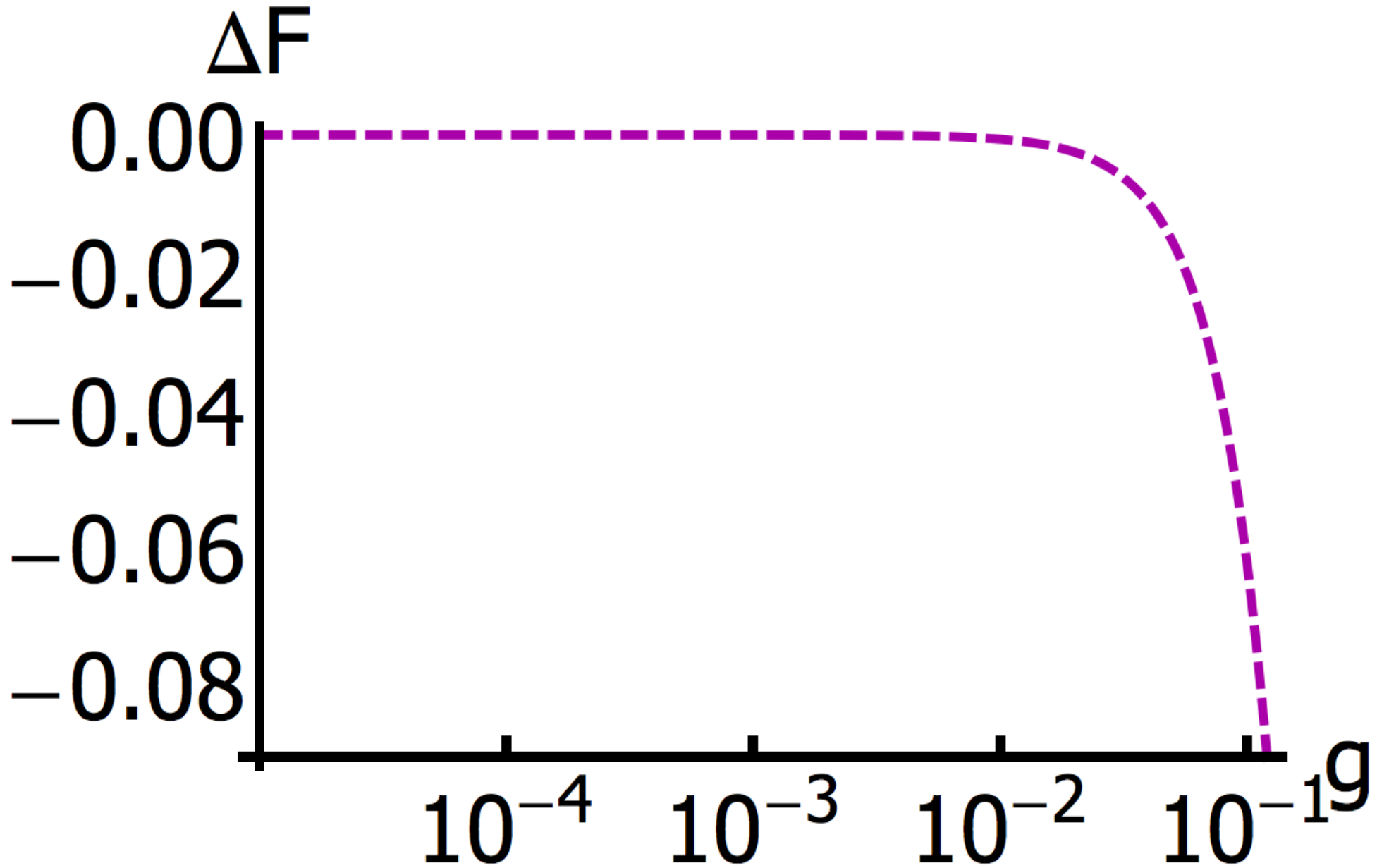}
\caption{Left: Log-linear plot of the free energy difference $\Delta F$ (in units of $\hbar \omega_\mathrm{m}$) as a function of the dimensionless temperature $\beta\hbar \omega_\mathrm{m}$
for $\omega_\mathrm{c}=500\omega_\mathrm{m}$, and $g=0.5\omega_\mathrm{m}$. Right: Log-linear plot of $\Delta F$  as a function of the scaled coupling $g/\omega_\mathrm{m}$ for 
$\omega_\mathrm{c}=500\omega_\mathrm{m}$, and $\beta=10^{-3}/\hbar\omega_\mathrm{m}$.\label{f:DeltaF}}
\end{figure}
\par
Using \eref{bch} we can easily compute the evolution of the initial Gibbs state, as defined by $\hat \varrho(t)=e^{-\frac{i}{\hbar}t\hat H_\mathrm{F}}\hat \varrho_\beta^\mathrm{(c)} 
\otimes \hat \varrho_\beta^\mathrm{(m)}e^{\frac{i}{\hbar}t\hat H_\mathrm{F}}$. In our case, it is easily seen that this always leads to a separable state, where any correlations between the 
optical and mechanical modes are fully classical. The dynamics is periodic in time: At $t=2\pi r/\omega_\mathrm{m}$ ($r\in\mathbb{Z}$), the system goes back to the initially factorized 
state, while for $t=(2r+1)\pi/\omega_\mathrm{m}$ ($r\in\mathbb{Z}$), one gets the maximally (classically) correlated state.
\par
Eq.~(\ref{bch}) also allows us to compute the partition function of the system, via a suitable Wick rotation of the argument, i.e., $u\to-i\hbar\beta$, which effectively identifies the imaginary 
time as an inverse temperature. For the initial state of the system the partition function factorizes in two canonical contributions $\mathcal{Z}_\mathrm{I}=\mathcal{Z}_{\beta}^\mathrm{(c)} 
\mathcal{Z}_{\beta}^\mathrm{(m)}\equiv[(1-e^{-\hbar\beta\omega_\mathrm{c}}) (1-e^{-\hbar\beta\omega_\mathrm{m}})]^{-1}$, while for the coupled system we obtain
\begin{equation}
\mathcal{Z}_\mathrm{F}=(1-e^{-\hbar\beta\omega_\mathrm{m}})^{-1} \sum_{n=0}^{\infty}e^{-\hbar\beta\omega_\mathrm{c}n}e^{\hbar\beta(g^2/\omega_\mathrm{m})n^2} \, .
\end{equation}%
The free energy difference is correspondingly given by
\begin{align}\label{DeltaF}
\Delta F=&-\frac{1}{\beta}\ln \left[\sum_{n=0}^{\infty}\frac{N_\mathrm{c}^n}{(1+N_\mathrm{c})^{n+1}}e^{\hbar\beta(g^2/\omega_\mathrm{m})n^2}\right] \nonumber \\
=&-\frac{1}{\beta}\ln \left[1-e^{-\hbar\beta\omega_\mathrm{c}}\right] \nonumber -\frac{1}{\beta} \ln \Biggl[\sum_{n=0}^{\infty}e^{-\hbar\beta\omega_\mathrm{c} n}e^{\hbar\beta\frac{g^2n^2}
{\omega_\mathrm{m}}}\Biggr],
\end{align}
which, as can be verified, agrees with the Jarzynski equality $\Delta F=-\frac{1}{\beta}\ln \chi(i\beta)$. Upon close inspection, it is readily apparent that the series involved in the latter 
expression is actually divergent. Indeed, for every finite value of $\beta$, $g/\omega_\mathrm{m}$, and $\omega_\mathrm{c}/\omega_\mathrm{m}$, there exists $\bar{n}=\bar{n}(g,r)$ 
such that $\forall n>\bar{n}$, we have that $g^2 n>r$. This causes the sum to diverge exponentially, such that $\Delta F$ is formally undefined. This divergent term can be traced 
back to the part of $\hat{H}_\mathrm{F}$ that reads $\omega_\mathrm{c}\hat{a}^{\dagger}\hat{a} -g^2/\omega_\mathrm{m}(\hat{a}^{\dagger}\hat{a})^2$. As is apparent, the spectrum of 
this Hamiltonian is not bounded from below. Occupation of levels with $n\geq\bar{n}$, which occurs naturally for any non-zero $\beta$, can thus be mapped into a negative temperature 
with respect to $\hat H_\mathrm{F}$. To resolve this issue, we impose a cutoff on the number of terms in the series; 
When $g/\omega_\mathrm{m}$ approaches or even exceeds unity, with the system entering the interesting strong-coupling regime of optomechanics, we must truncate the series to 
correspondingly small photon numbers in order to prevent dynamical instability, and the ensuing divergence of $\Delta F$, upon quenching the system. For the rest of this work, we will 
therefore restrict ourselves to the physical domain in which the series does converge.  
\par
An explicit calculation of $\Delta F$, as illustrated in Fig.~\ref{f:DeltaF}, shows that the free energy difference is negative, in agreement with the statement of the second law $\Delta 
F\le \langle W\rangle \equiv 0$. Moreover, the irreversible work reduces to $W_\mathrm{irr}=- \Delta F$. Upon moving towards lower temperatures, both the evolved state and the reference 
thermal state tend to collapse onto the ground state, leading to vanishing values of the irreversible work, as is apparent from the figure. On the other hand, upon increasing the coupling 
$g/\omega_\mathrm{m}$, the free energy difference grows in modulus. 

\subsection{Initial displacement of the mechanical oscillator}
In the previous subsection we observed how $\langle W\rangle=0$ for an initial thermal state of the Hamiltonian $H_\mathrm{I}$, independently of the strength of the quench. The fact can be 
seen as a direct consequence of the symmetry of the interaction which, being proportional to $\hat{x}$, is an odd function in the mechanical field operators, 
such that
\begin{equation}
\langle W\rangle
=-g\,N_\mathrm{c}\tr{( \hat{b}+\hat{b}^{\dagger})\hat \varrho_\beta^\mathrm{(m)}}=0.
\end{equation}
In other words, the average work generated by this kind of quench will be zero. In order to remedy this, we now add an initial displacement of amplitude $\mathcal{E}\,\omega_\mathrm{m}
\in{\mathbb R}$ to the mechanical mode $\hat{b}$ of the Hamiltonian~(\ref{Hgen}) so that the initial and final Hamiltonians will now  read $\hat H_{\mathrm{I,F},\mathcal{E}}=\hat H_\mathrm{I,F} 
+ \hbar\, \mathcal{E}\,\omega_\mathrm{m}(\hat{b}+\hat{b}^{\dagger})$.  
It can be shown that $\hat H_{\mathrm{I},\mathcal{E}}=\hat D(\mathcal{E})\hat H_\mathrm{I} \hat D^{\dagger}(\mathcal{E})$ and $\hat H_{\mathrm{F},\mathcal{E}}=\hat D(\mathcal{E})
(\hat H_\mathrm{F}+2\hbar\, g\,\mathcal{E}\,\hat{a}^{\dagger}\hat{a} )\hat D^{\dagger}(\mathcal{E})$ with $\hat D({\cal E})$ a local displacement of amplitude $\mathcal{E}$. Proceeding as before, 
the characteristic function of the work distribution can be computed as
\begin{multline}\label{chidisp}
\chi(u)=\sum_{n=0}^{\infty}\frac{N_\mathrm{c}^n}{(1+N_\mathrm{c})^{n+1}}e^{-i(g/\omega_\mathrm{m})^2n^2(\omega_\mathrm{m}u-\sin\omega_\mathrm{m}u)}\\
\times e^{-(g/\omega_\mathrm{m})^2n^2(1+2 N_\mathrm{m})(1-\cos\omega_\mathrm{m}u)}e^{2ign\mathcal{E}u} \,,
\end{multline}
which differs from \eref{chi} by a phase factor. This extra factor is actually responsible for positive derivative of the imaginary part $\mathrm{Im}[\chi(u,\mathcal{E})]$ at the origin 
and hence to a non-zero value of the average work. Indeed, applying \eref{avwork}, one finds that the average work done by quenching the optomechanical interaction is given by
\begin{equation}\label{work}
\langle W \rangle=2\hbar\,g\,\mathcal{E}N_\mathrm{c} \, ,
\end{equation}
which {depends linearly on} the displacement $\mathcal{E}$, {on} the number of thermal photons populating the cavity, and on the quenching parameter.  
\par
\begin{figure}[t] 
\centering 
\includegraphics[scale=.21]{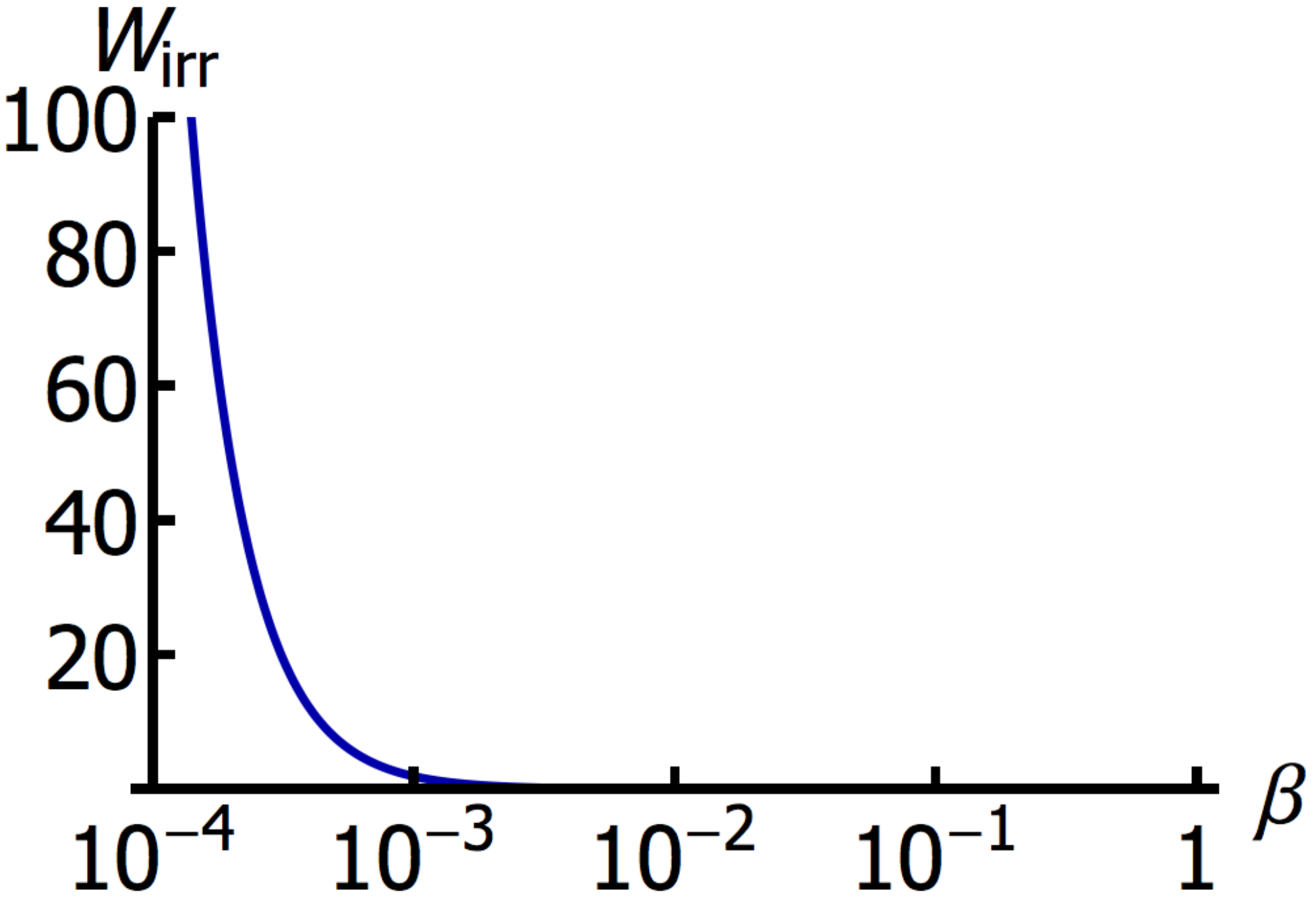}
\includegraphics[scale=.21]{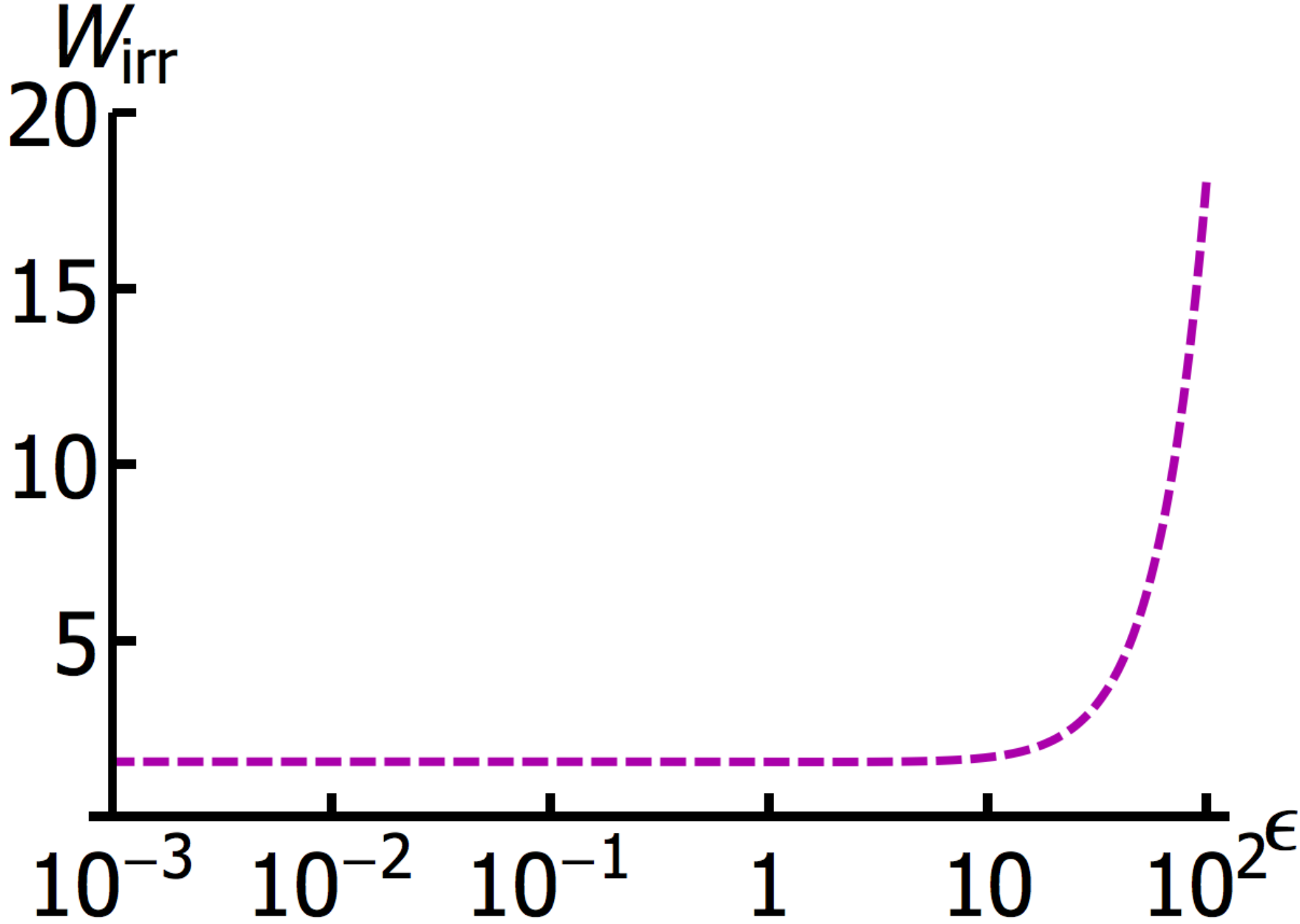}
\caption{Left: Log-linear plot of the irreversible work $W_\mathrm{irr}$ (in units of $\hbar \omega_\mathrm{m}$) as a function of the dimensionless temperature $\beta\hbar \omega_\mathrm{m}$
for $\omega_\mathrm{c}=500\omega_\mathrm{m}$, and $g=0.5\omega_\mathrm{m}$. Right: Log-linear plot of $W_\mathrm{irr}$ as a function of the mechanical displacement $\mathcal{E}$ for 
$\omega_\mathrm{c}=500\omega_\mathrm{m}$, and $\beta=10^{-3}/\hbar\omega_\mathrm{m}$.\label{f:WirrLin}}
\end{figure}

Finally, the free energy difference for this model is given by
\begin{equation}
\Delta F=-\frac{1}{\beta}\ln \left[\sum_{n=0}^{\infty}\frac{N_\mathrm{c}^n}{(1+N_\mathrm{c})^{n+1}}e^{\hbar\beta\frac{g^2n^2}{\omega_\mathrm{m}}-2\hbar\beta g n \mathcal{E}}\right].
\end{equation}
The behavior of the irreversible work $W_\mathrm{irr}$ is reported in Fig.~\ref{f:WirrLin}, with respect to the inverse temperature and the magnitude of the displacement.

\par

\subsection{Quenched quadratic optomechanical interaction}
\begin{figure}[t] 
\centering 
\includegraphics[scale=.46]{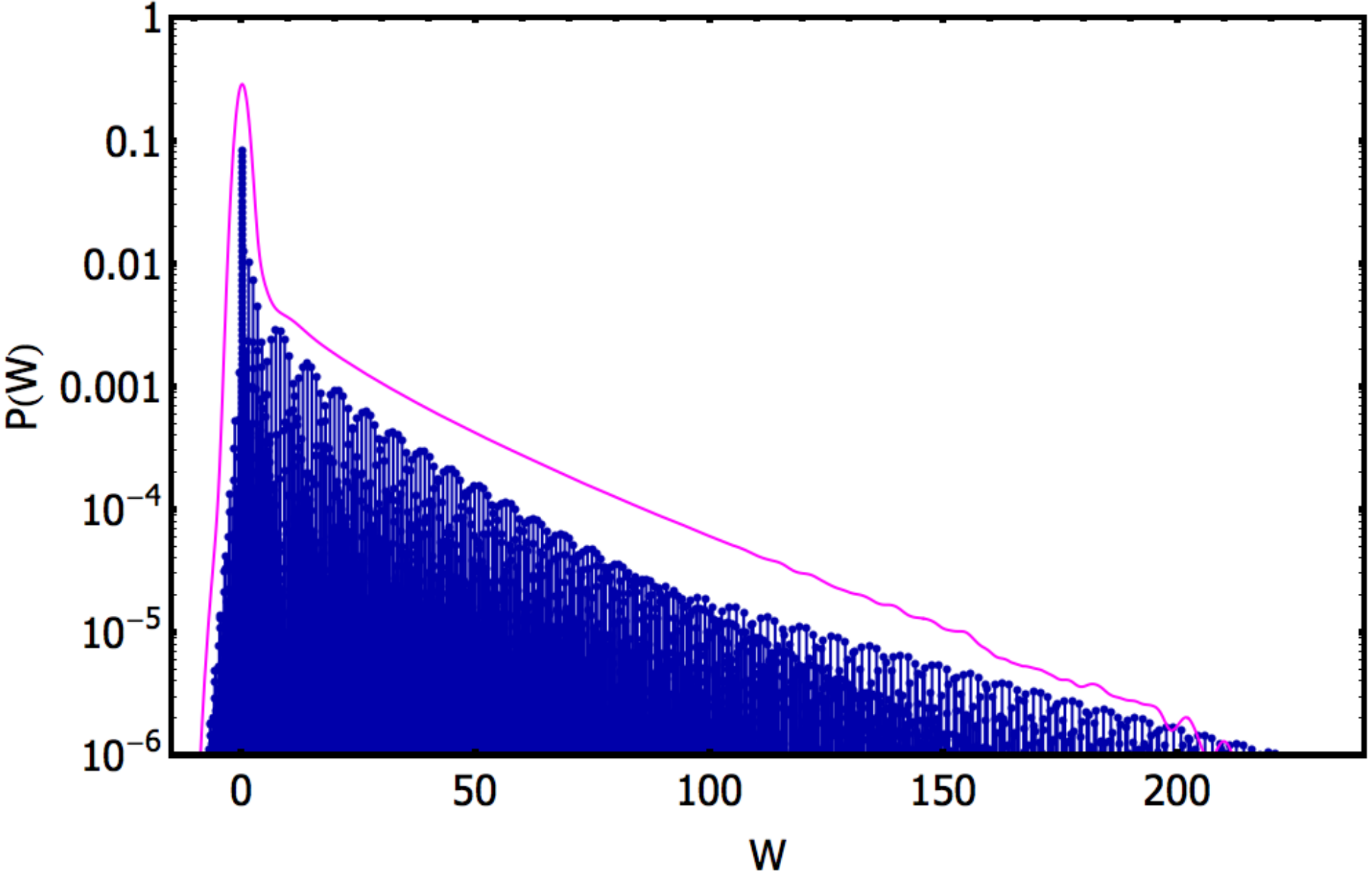}
\caption{Logarithmic plot of the probability distribution of work (in units of $\hbar \omega_\mathrm{m}$) corresponding to the parameters ($N_\mathrm{c}, N_{\rm m}, \kappa)
=(0.19,9,0.7\omega_\mathrm{m}$). We also show the coarse grained version of the work distribution (solid magenta line). The coarse graining is realized by convolving the 
discrete distribution with a Gaussian function of standard deviation $0.9\hbar\omega_\mathrm{m}$.
\label{f:CGquadratic}}
\end{figure}

\begin{figure}[h] 
\centering 
\includegraphics[scale=.21]{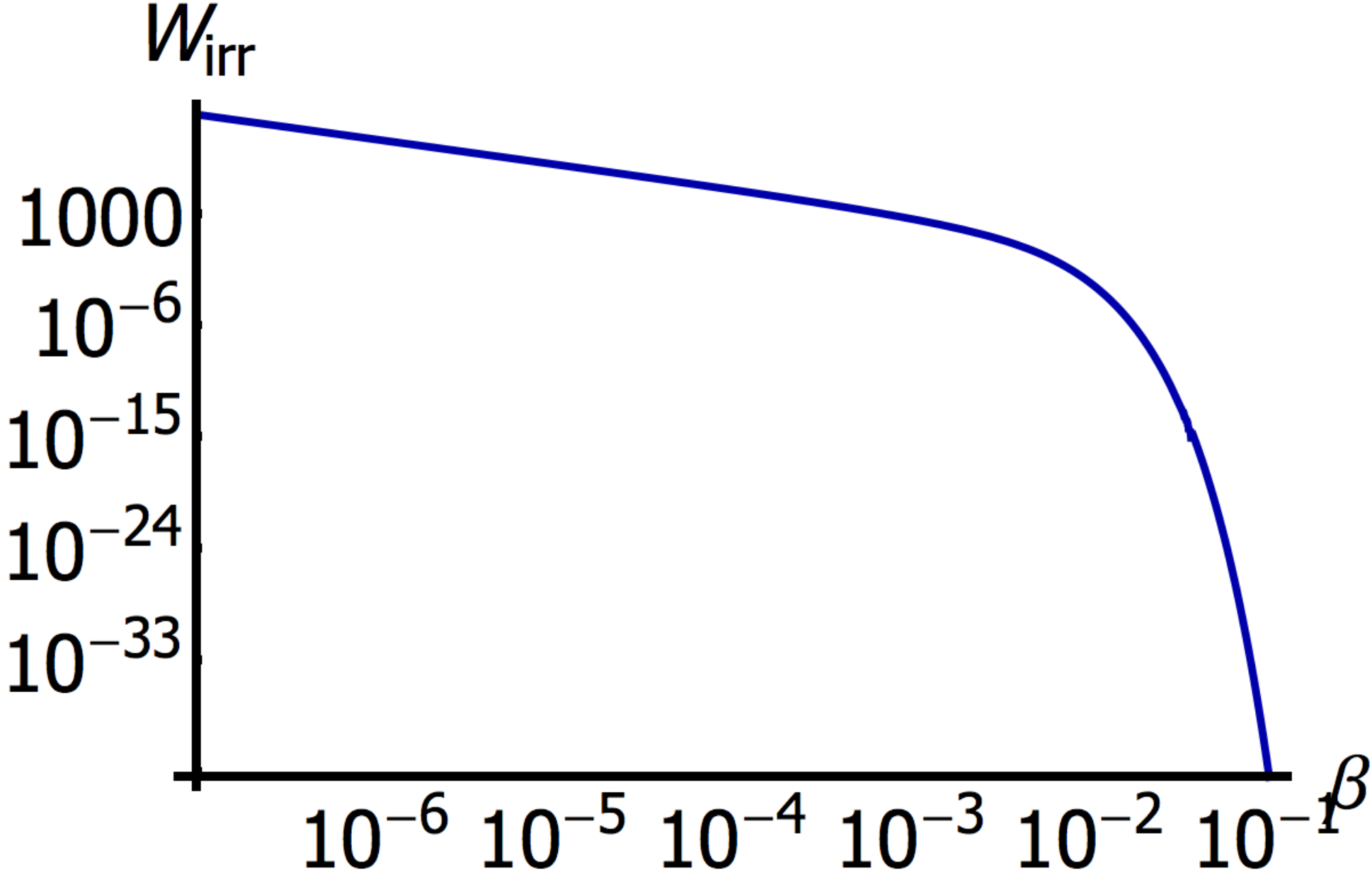}
\includegraphics[scale=.21]{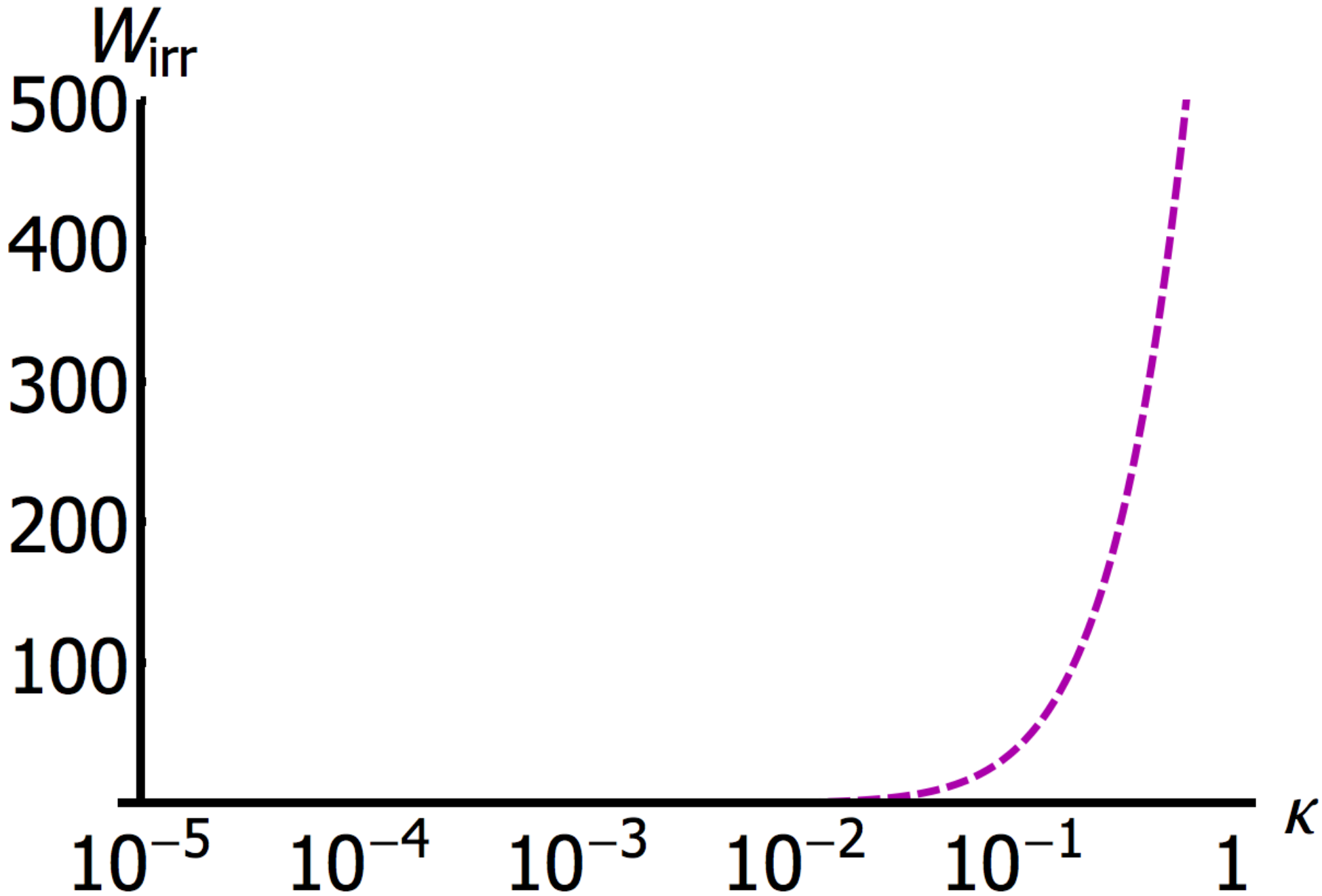}
\caption{Left: Log-log plot of the irreversible work $W_\mathrm{irr}$ (in units of $\hbar \omega_\mathrm{m}$) as a function of the dimensionless temperature $\beta\hbar \omega_\mathrm{m}$
for $\omega_\mathrm{c}=10^3\omega_\mathrm{m}$, and $g=0.5\omega_\mathrm{m}$. Right: Log-linear plot of $W_\mathrm{irr}$ as a function of the coupling strength $\kappa$ for 
$\omega_\mathrm{c}=10^3\omega_\mathrm{m}$, and $\beta=10^{-3}/\hbar\omega_\mathrm{m}$.\label{f:WirrQuad}}
\end{figure}

We will consider now the case where the photon number operator of the cavity field is coupled to the square of the position operator of the mirror. As before, we will
concentrate on the single-photon regime where the interaction of a single photon with the mechanical mode is enough to appreciably change its frequency 
and also squeeze its state. In this instance, we can introduce the single-photon coupling strength $\kappa$ through the relation $G^{(2)}=\kappa/x_\mathrm{zpf}^2$, in 
analogy with the linear case. The initial Hamiltonian $H_\mathrm{I}$ is  unmodified and still given by \eref{HI}, whereas the the post-quench Hamiltonian now reads
\begin{equation}  
\hat H_\mathrm{F}=\hat H_\mathrm{I}+\hbar \,\kappa\,\hat{a}^{\dagger}\hat{a}(\hat{b}+\hat{b}^{\dagger})^2.
\end{equation}
We choose to work with a non-negative $\kappa$, since $\kappa<0$ can introduce post-quench instabilities similar to the one noted for the linear case. The $\kappa>0$ 
case exhibits no such instabilities. Yet again, we see that this interaction preserves the photon number $\hat{a}^{\dagger}\hat{a}$, so that it proves convenient to write 
$\hat H_\mathrm{F}=\bigoplus_{n=0}^{\infty}\hat H_{\mathrm{F},n}$ {where} each $\hat H_{F,n}$ can be cast in the form
\begin{equation}\label{H_{F,n}}
\hat H_{\mathrm{F},n}=\Bigl[\hbar\omega_\mathrm{c}n+\hbar\Omega_n\bigl(\hat{b}^{\dagger}\hat{b}+\tfrac{1}{2}\bigr)
+\hbar\Sigma_n\bigl(\hat{b}^{\dagger\, 2}+\hat{b}^2\bigr)\Bigr]\ket{n}{\bra{n}}_{\rm c}\, ,
\end{equation}  
where $\Omega_n\equiv\omega_\mathrm{m}+2\,\kappa\,n$ and $\Sigma_n\equiv 2 \kappa\,n$. Within each such fixed photon-number manifold, we notice the appearance 
of a modified mechanical frequency, together with a squeezing operator for the mechanical mode whose argument is conditioned on the photon number. The evolution 
operator relative to the post-quench Hamiltonian can subsequently be expressed as
\begin{equation}
\label{quadratica}
e^{-\tfrac{i}{\hbar} \hat H_\mathrm{F} u}
=\sum_{n=0}^{\infty}e^{-iu[\omega_\mathrm{c}n+\Omega_n(\hat{b}^{\dagger}\hat{b}+\frac{1}{2})+\Sigma_n(\hat{b}^{\dagger\, 2}+\hat{b}^2)]}\ket{n}{\bra{n}}_{\rm c}\,.
\end{equation}
Our next task is to disentangle each exponential operator in the sum. 
By using the commutation relations between the operators involved in Eq.~(\ref{quadratica}), which provide a two-excitation realization of the $\mathfrak{su}(1,1)$ algebra~
\cite{Agarwal}, we find 
{\begin{multline}
e^{-\tfrac{i}{\hbar}\hat H_{\mathrm{F},n} u}=e^{\tfrac{1}{2}[\xi_n^*\hat{b}^2-\xi_n\hat{b}^{\dagger2}]}e^{-i\eta_n(\hat{b}^{\dagger}\hat{b}+\tfrac{1}{2})}e^{-i\omega_\mathrm{c}u\,n}
\ket{n}{\bra{n}}_\mathrm{c}\,,
\end{multline}}
where
\begin{equation}
\eta_n\equiv\mathrm{arctan}\left[\frac{1+2\tilde\kappa\,n}{\sqrt{1+4\tilde\kappa\,n}} \tan\bigl(\omega_\mathrm{m}u\sqrt{1+4\tilde\kappa\,n}\bigr) \right]
\end{equation}
with $\tilde\kappa\equiv\kappa/\omega_\mathrm{m}$ being a dimensionless quench parameter. We further have the complex quantity $\xi_n\equiv\vert\xi_n\vert e^{i\phi_n}$ whose 
phase is $\phi_n\equiv\eta_n+\frac{\pi}{2}$ and modulus
\begin{equation}
\vert\xi_n\vert\equiv\mathrm{arcsinh}\Biggl[\frac{2\tilde\kappa\,n}{\sqrt{1+4\tilde\kappa\,n}} \sin\bigl(\omega_\mathrm{m}u\sqrt{1+4\tilde\kappa\,n}\bigr)\Biggr] \, .
\end{equation}%
Armed with this tool we can thus compute the characteristic function of the work distribution, which reads
\begin{equation}\label{chiq}
\chi(u)=\sum_{n=0}^{\infty}\frac{N_\mathrm{c}^n}{(1+N_\mathrm{c})^{n+1}}\frac{1}{\sqrt{\sum^2_{j=0}\chi_{n,j}N^j_{\rm m}}}\,,
\end{equation}%
and comes in the form of a thermal average with respect to the cavity distribution---as in \eref{chi}---of algebraic functions. Each of the latter is the reciprocal of the square-root of 
a second degree polynomial in the mean number of phonons $N_\mathrm{m}$, whose coefficients are concisely related to each other. Indeed, we can split $\chi_{n,0}$
into its real and imaginary parts, which read
\begin{figure}[t] 
\centering
\includegraphics[scale=.29]{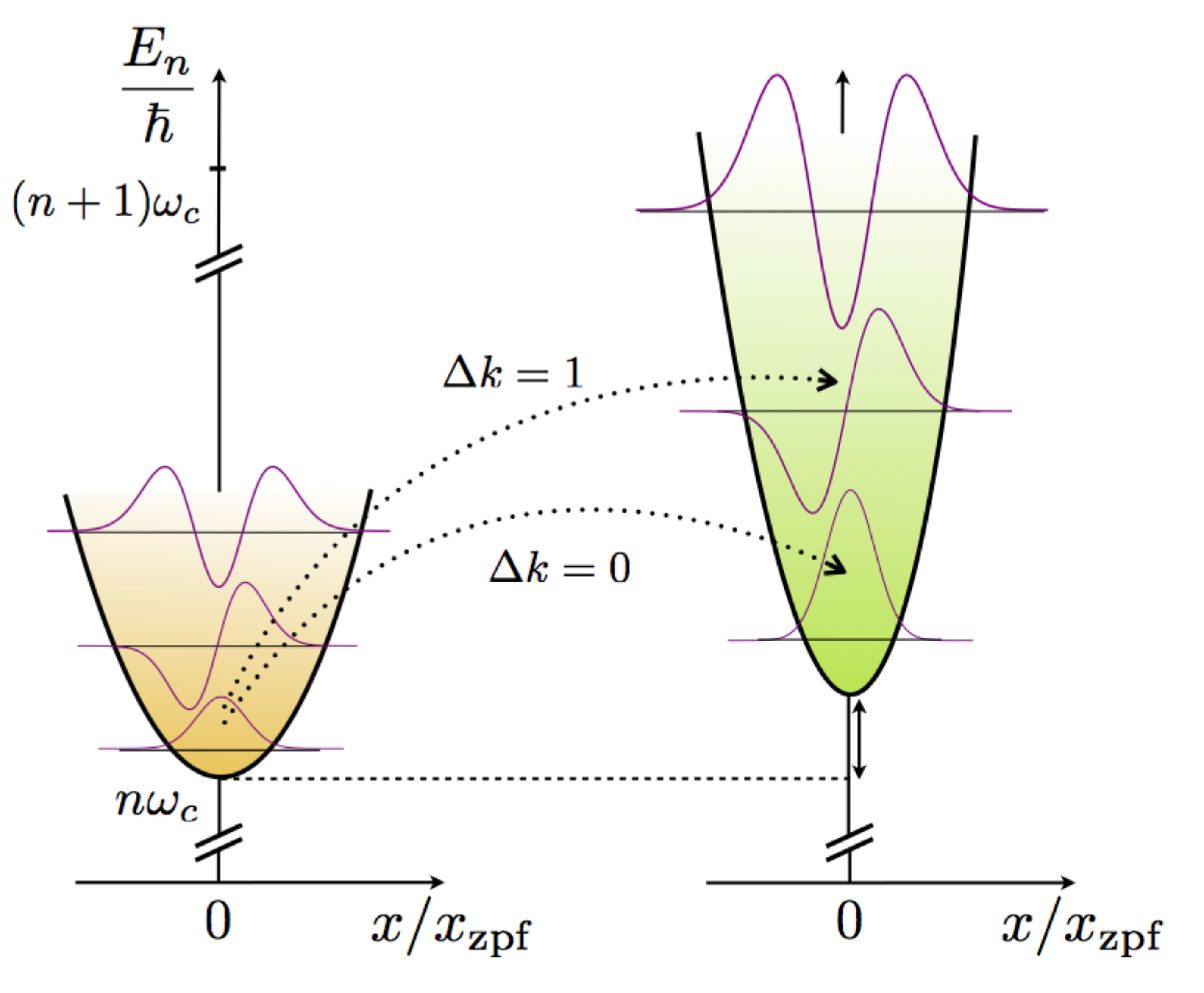}
\caption{Schematic diagram (not to scale) of the energy-level structure of the pre-quench, $\hat H_{\mathrm{I},n}$, and post-quench, $\hat H_{F,n}$, Hamiltonians for the $n$-photon 
manifold. Quenching the quadratic optomechanical interaction results both in an energy shift and a squeezing of the frequency of the machanical oscillator. Two possible transitions 
induced by the quench---having different values of $\Delta k=k'-k$---are shown as an example.\label{f:quadquench}}
\end{figure}
\begin{equation}
\begin{aligned}
\mathrm{Re}(\chi_{n,0})&= \cos(\omega_\mathrm{m}u)\cos(\omega_\mathrm{m}u\sqrt{1+4\tilde\kappa\,n})\\
&+\frac{1+2\tilde\kappa\,n}{\sqrt{1+4\tilde\kappa\,n}}\sin(\omega_\mathrm{m}u)\sin\bigl(\omega_\mathrm{m}u\sqrt{1+4\tilde\kappa\,n}\bigr) \, ,
\end{aligned}
\end{equation}
and
\begin{equation}
\begin{aligned} 
\mathrm{Im}(\chi_{n,0})&= \sin(\omega_\mathrm{m}u)\cos\bigl(\omega_\mathrm{m}u\sqrt{1+4\tilde\kappa\,n}\bigr)\\
&-\frac{1+2\tilde\kappa\,n}{\sqrt{1+4\tilde\kappa\,n}}\cos(\omega_\mathrm{m}u)\sin\bigl(\omega_\mathrm{m}u\sqrt{1+4\tilde\kappa\,n}\bigr).
\end{aligned}
\end{equation}%
We thus have $\chi_{n,1}= 2(\chi_{n,0}-1)\ \mathrm{and}\ \chi_{n,2}= 2\bigl [\mathrm{Re}(\chi_{n,0})-1\bigr]$. As before, since the Fourier transform of \eref{chiq} cannot be directly evaluated, 
in order to compute the probability distribution of the work \eref{pdf} we proceed by diagonalizing the post-quench Hamiltonian $\hat H_\mathrm{F}$. First, we keep in mind that $\hat H_\mathrm{I}$ 
is the same as before. However, within any fixed photon number manifold, $\hat H_{\mathrm{F},n}$ be diagonalized via a squeezing operation $\hat S(z)=\exp(z^* \hat{b}^2/2-z\, \hat{b}
^{\dagger\, 2}/2)$ on the mechanical mode conditioned on the photon number $n$~\cite{Nori}. Once again denoting the post-quench quantities with a prime, and expressing the states in the eigenbasis of 
$\hat H_\mathrm{I}$, we find eigenstates $\hat H_{\mathrm{F},n}\ket{n'}_\mathrm{c}  \otimes \hat S(\zeta_{n^\prime}) \ket{k'}_\mathrm{m} =E_{n',k'}\ket{n'}_\mathrm{c}  \otimes \hat S(\zeta_{n^\prime}) 
\ket{k'}_\mathrm{m}$, where the squeezing parameter is given by $\zeta_{n^\prime}\equiv\tfrac{1}{4}\log\bigl[1+4(\kappa/\omega_\mathrm{m})\,{n^\prime}\bigr]$, and the eigenvalue
\begin{equation}
E_{n',k'}=\hbar\omega_\mathrm{c}\,n'+\hbar\omega_\mathrm{m}\sqrt{1+4(\kappa/\omega_\mathrm{m})\,n^\prime}\,(k'+\tfrac{1}{2})\,.
\end{equation}%
As sketched in Fig.~\ref{f:quadquench}, for the manifold corresponding to $n^\prime$ photons, the quench results in a modification of the oscillation frequency which, is multiplied by a factor
$\sqrt{1+4(\kappa/\omega_\mathrm{m})\,n^\prime}$, a relative shift of the mechanical levels by $\hbar\omega_\mathrm{m}\bigl[\sqrt{1+4(\kappa/\omega_\mathrm{m})\,n^\prime}-1\bigr]$,  
and a squeezing of the state by a factor $\zeta_{n^\prime}$. 
Putting everything together, the probability distribution of the work is thus given by
\begin{equation}
\begin{aligned}
\label{PWq}
P(W)&= \sum_{n,n',k,k'}p_n^\mathrm{(c)}p_k^\mathrm{(m)}\left\vert \bra{k'}S(\zeta_{n'})\ket{k}\right\vert^2 \\
&\times\,\delta\left[W-(E_{n',k'}-E_{n,k})\right] \delta_{n,n'}\\
&= \sum_{n,k,k'}p_n^\mathrm{(c)}p_k^\mathrm{(m)} \frac{k!\, k'!}{(\cosh \zeta_n)^{2k+1}}\left[\mathcal{S}(k,k',\zeta_n)\right]^2 \\
&\times \delta\bigl\{W-\hbar\omega_\mathrm{m}\bigl[\sqrt{1+\frac{4n\kappa}{\omega_\mathrm{m}}}\,k'-k\bigr]\bigr\}\, ,
\end{aligned}
\end{equation}%
where $\mathcal{S}(k,k',\zeta_n)$ is given by
\begin{equation}
\begin{aligned}
\mathcal{S}(k,k',\zeta_n)&=\sum_{m=0}^{\lfloor \frac{k'}{2}\rfloor}\sum_{l=0}^{\lfloor \frac{k}{2}\rfloor} \frac{(-1)^{3m+2l}}{2^{m+l}\,m!\,l!} \frac{(
\tanh\zeta_n)^{m+l}}{(k-2l)!} \\
&\times(\cosh\zeta_n)^{2l} \delta_{k'-2m,k-2l} \, ,
\end{aligned}
\end{equation}
being $\lfloor x\rfloor$ the floor function of argument $x$, which yields the largest integer not greater than $x$.
\begin{figure}[t] 
\centering 
\includegraphics[scale=.42]{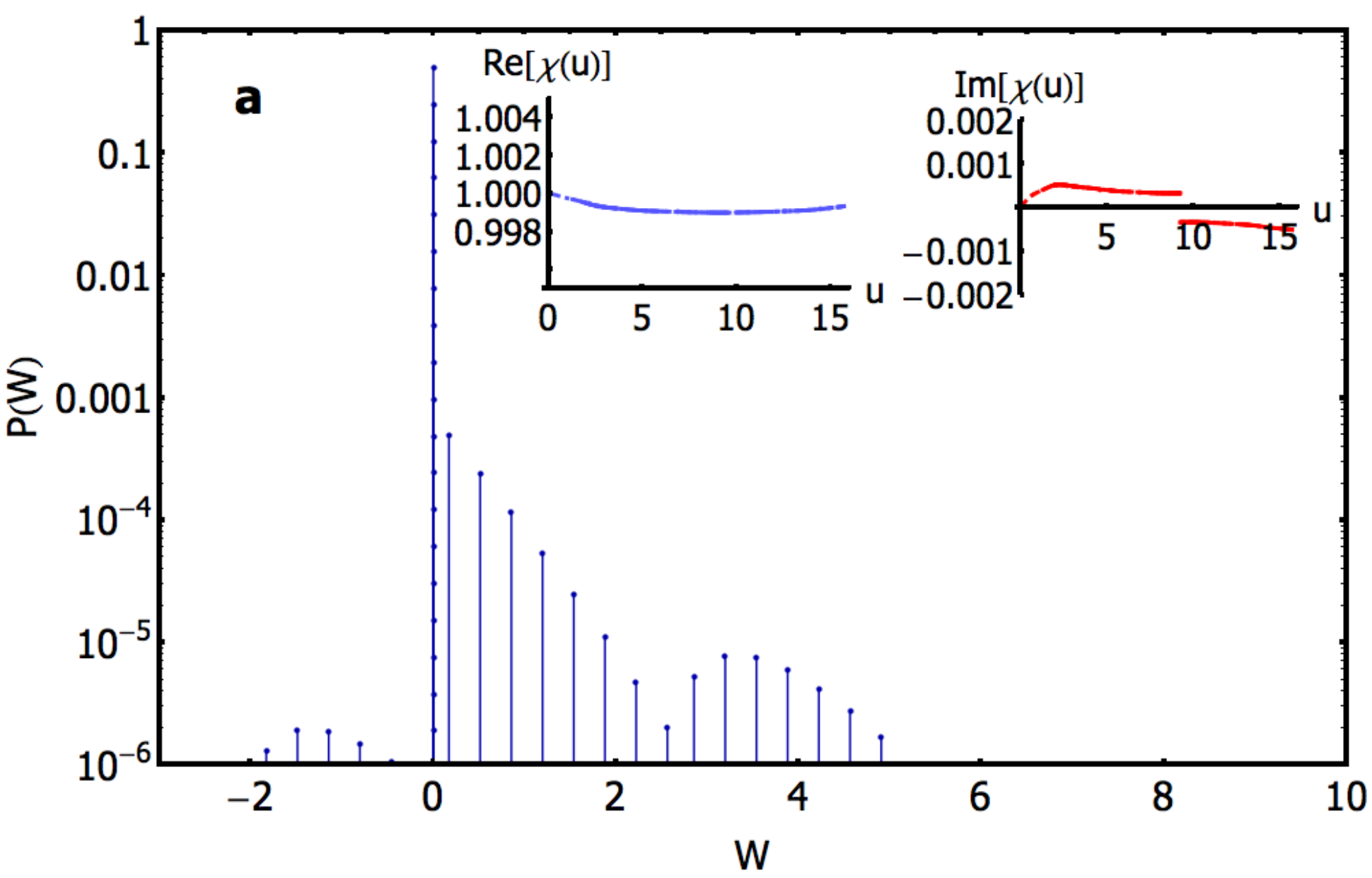}
\\
\includegraphics[scale=.42]{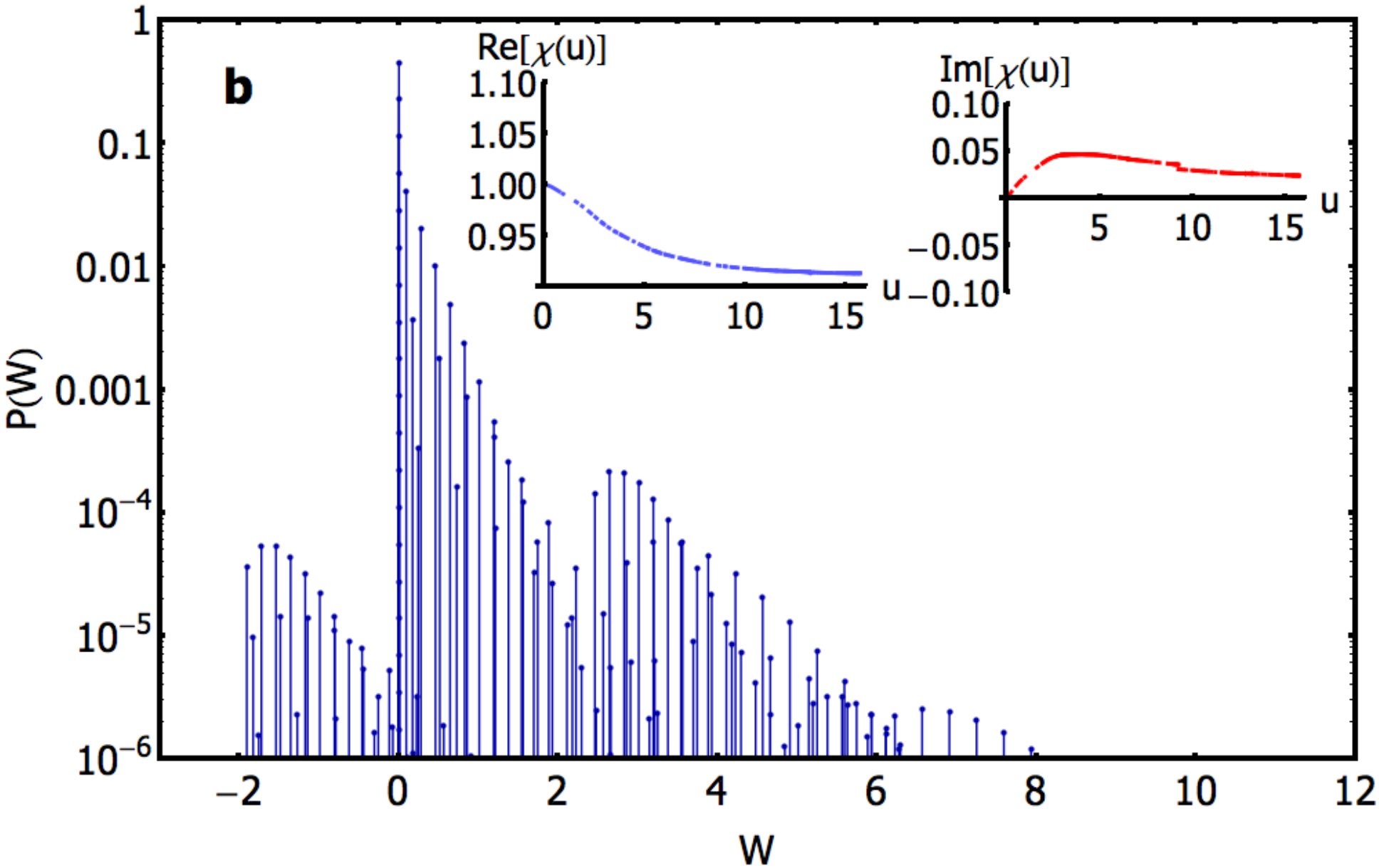}
\\
\includegraphics[scale=.42]{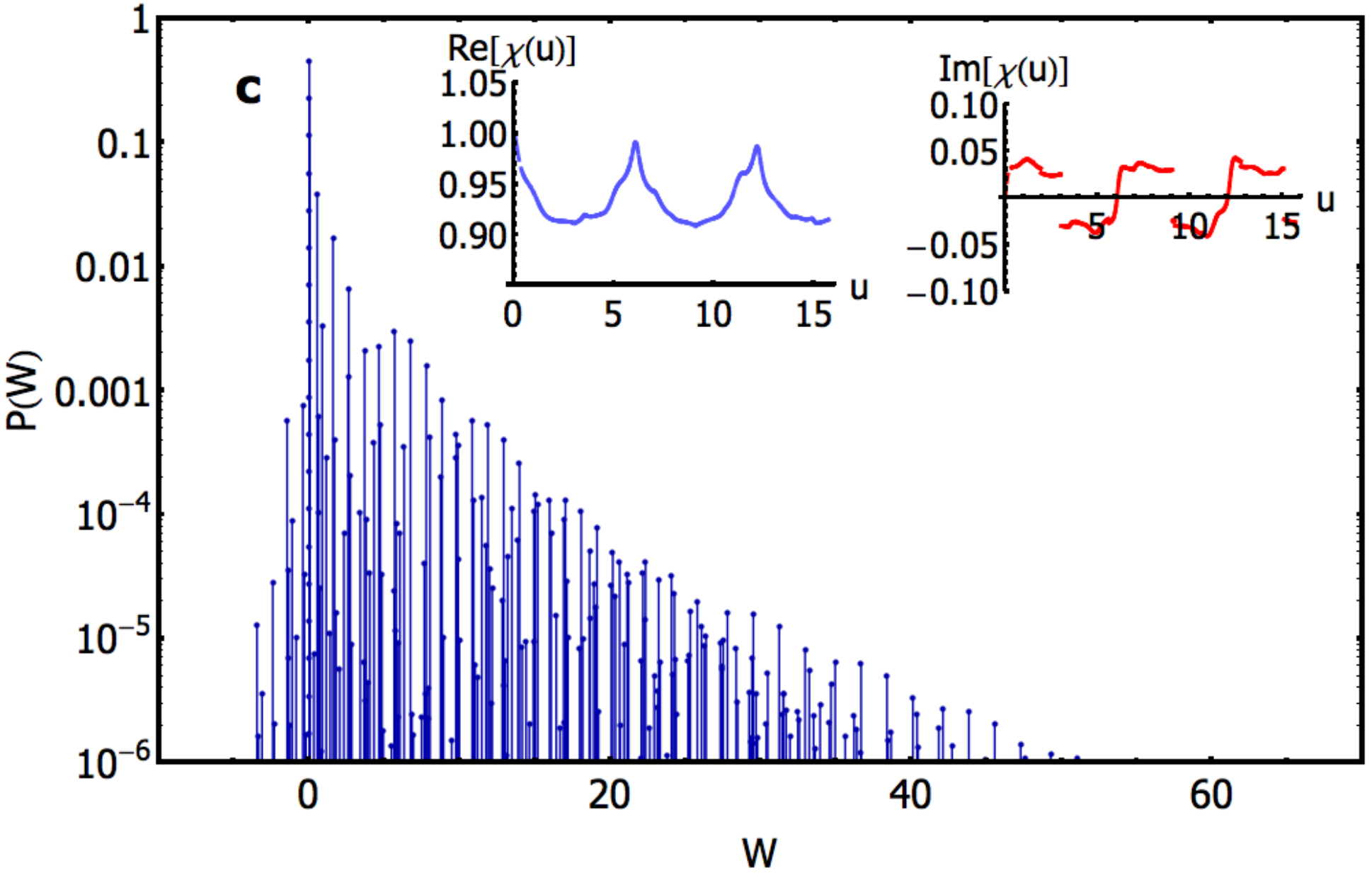}
\caption{Logarithmic plot of the probability distribution of the stochastic work variable, $W$ (in units of $\hbar \omega_\mathrm{m}$) for different values of the average number of 
cavity photons $N_\mathrm{c}$, average number of mechanical phonons $N_\mathrm{m}$ and coupling $\kappa$. Panel ({\bf a}) is for $(N_{\mathrm{c}}, N_{\rm m}, \kappa)=(0.001,0.1,0.2
\omega_\mathrm{m})$, ({\bf b}) is for $(N_{\mathrm{c}}, N_{\rm m}, \kappa)=(0.1,1,0.1\omega_\mathrm{m})$ while ({\bf c}) for $(N_{\mathrm{c}}, N_{\rm m}, \kappa)=(0.1,1,0.8\omega_\mathrm{m})$.  
In the inset is shown the behavior against the time-like variable $u$ (multiplied by $\omega_\mathrm{m}$) of the real, $\mathrm{Re}(\chi)$ (solid blue, left), and imaginary, 
$\mathrm{Im}(\chi)$ (dashed red, right), parts of the characteristic function.\label{f:Chiq}}
\end{figure}%
The probability distribution for the work done on the oscillator in the case of a quadratic interaction, as derived in this section, is illustrated for some representative cases in Figs.~\ref{f:Chiq} 
and~\ref{f:CGquadratic}. In order to characterize quantitatively the key features of the distribution of work, here we mention that the average work generated by a quench of the quadratic 
optomechanical Hamiltonian is different from zero and is then given by
\begin{equation}
\langle W \rangle=\hbar\kappa N_\mathrm{c} (1+2N_\mathrm{m}),
\end{equation}%
hence increasing with respect the occupation numbers of both the cavity and the mechanical mode, as made apparent by inspecting the different panels in Fig.~\ref{f:Chiq}. The variance 
of the distribution reads
\begin{equation}
\langle W^2 \rangle-\langle W \rangle^2=\hbar^2\kappa^2N_\mathrm{c} (3+5N_\mathrm{c})(1+2N_\mathrm{m})^2 \,.
\end{equation}
Finally, the most striking feature of the probability distribution in the case of a quadratic quench is that it is very asymmetrical, fact witnessed by its skewness 
\begin{equation}
\gamma=\frac{4+8N_\mathrm{c}+(g/\omega_\mathrm{m})(15+81N_\mathrm{c}+74N_\mathrm{c}^2)(1+2N_\mathrm{m})^2}{(g/\omega_\mathrm{m})\sqrt{N_\mathrm{c}}(3+5N_\mathrm{c})^{3/2}
(1+2N_\mathrm{m})^2}.
\end{equation}
We note that, for $N_\mathrm{m}\gg1$, it acquires the values $5/\sqrt{3N_\mathrm{c}}$ for $N_\mathrm{c}\ll1$ and $74/5\sqrt{5}$ for $N_\mathrm{c}\gg1$; both these values are 
independent of the strength of the quench. 
As for the linear case the dynamics brings the initial bipartite state of cavity and mechanical mode into a separable sate, given by $\hat \varrho(t)=e^{-\frac{i}{\hbar}t\hat H_\mathrm{F}}\hat 
\varrho_\beta^\mathrm{(c)} \otimes \hat \varrho_\beta^\mathrm{(m)}e^{\frac{i}{\hbar}t\hat H_\mathrm{F}}=\sum_n p_n^\mathrm{(c)} \ket{n}{\bra{n}}_\mathrm{c}\otimes \int\mathrm{d}^2
\alpha\, {\cal P}^\mathrm{(m)}(\alpha)\ket{e^{i\eta_n}\alpha,\xi_n}\bra{e^{i\eta_n}\alpha,\xi_n}_\mathrm{m}$ where $\ket{e^{i\eta_n}\alpha,\xi_n}_\mathrm{m}=\hat S(\xi_n) \hat 
D(e^{i\eta_n}\alpha) \ket{0}_\mathrm{m}$ is a squeezed coherent state of the mechanical mode, and hence no entanglement is generated between the two modes.
Proceeding in the same manner as before, we can show that the free energy can be cast in the form
\begin{equation}
\begin{aligned}
\Delta F&= -\frac{1}{\beta}\ln\bigl[\sinh\bigl(\tfrac{\beta}{2}\bigr)\bigr]\\
&-\frac{1}{\beta}\ln\Biggl[\sum_{n=0}^{\infty}\frac{N_\mathrm{c}^n\,\mathrm{cosech}\bigl(\sqrt{1+\tfrac{4n\kappa}{\omega_\mathrm{m}}}\,\tfrac{\beta}{2}\bigr)}{(1+N_\mathrm{c})^{n+1}}\Biggr].
\end{aligned}
\end{equation}
In this case, too, a suitable Wick-like rotation to imaginary $u$ can be performed to obtain $\Delta F$ from $\chi(u)$. In practice, however, this calculation is frought with technical difficulties and 
it is far easier to compute $\Delta F$ from an explicit diagonalisation of the Hamiltonian, as was done above. The behavior of the irreversible work for this case has been shown in Fig. (\ref{f:WirrQuad}),
and once again we can see how it drops lowering the temperature and increases by increasing the coupling strength.
\par
As in the linear case, is easier to extract a physical meaning behind the various features of these plots by inspecting the respective coarse-grained distributions. 
First, we see that the positive-$W$ tail still exhibits an approximately exponential decay. It is also apparent that the distribution is, in this case, significantly more skewed towards the right 
than in the linear case, which can be understood simply through the fact that the post-quench mechanical oscillator frequency is always \emph{larger}; even for the case when $k^\prime=k$, 
therefore, which at least for small $\kappa/\omega_\mathrm{m}$ has a large probability of occurring, the work done is positive.

\par

\section{Conclusions and outlook}
\label{conc}
The exploration of out-of-equilibrium features of small systems working in the quantum regime is attracting ever-increasing attention. Optomechanical systems, more so than other systems, 
offer the tantalizing perspective of naturally bridging the study of quantum thermodynamics with the macroscopic domain. We actually believe that this class of systems offers the possibility of 
a captivating analogy: Movable mirrors and cavity fields closely resemble pistons and working media in a piston--chamber engine; in turn, this embodies the archetypal example of a thermal 
machine. In this sense, such systems may serve as the paradigm for understanding a new class of machines, operating  both in the quantum regime and far from equilibrium. 
However, an adequate description of optomechanical systems involves a fully quantum treatment, and a detailed analysis of the thermodynamical properties of them, carried out at a fundamental 
level and retaining the full nonlinearity of the interaction, has not been conducted thus far. In this work we discussed the generation of work induced by a non-equilibrium transformation in an 
isolated optomechanical system, quantitatively assessing how an instantaneous quench of the light--matter coupling affects the thermodynamical response of the system. Our study was grounded 
through several analytic results, presenting expressions for both the characteristic function of the work distribution and the full statistics of the work generated for two different situations of much 
relevance for current and future optomechanical experiments. For a quench of linear coupling between light and the position of an oscillator, we found that no work is generated on average, whilst 
quenching a quadratically-coupled optomechanical interaction requires work to be performed on the system.
\par
Besides being interesting in itself, and allowing for a full analytical treatment, the scenario we addressed comprises the fundamental ingredients necessary in order to gain knowledge about the 
microscopic origin of the work generated by quenching an optomechanical interaction, from a fully quantum perspective. An in-depth understanding of the thermodynamical response of such an 
isolated quantum system represents the cornerstone for future investigations. For instance, the implementation of protocols for extracting work out of such systems will require benchmarks based 
on the analysis that we have performed here, which will in turn be necessary to help uncover fundamental advantages or limitations for possible future thermal machines working in the quantum 
regime and that exploit the optomechanical interaction. 
\par
\section*{Acknowledgments}
We are grateful to M. Aspelmeyer for discussions and encouragements. This work was supported by the UK EPSRC (EP/L005026/1 and EP/J009776/1), the John Templeton Foundation 
(grant ID 43467), the EU Collaborative Project TherMiQ (Grant Agreement 618074), and the Royal Commission for the Exhibition of 1851. Part of this work was supported by COST Action 
MP1209 ``Thermodynamics in the quantum regime''.


\end{document}